\documentclass[a4paper,12pt,reqno]{amsart}
\usepackage{latexsym}
\usepackage{amssymb,color}

\newtheorem{thm}{Theorem}[section]

\newtheorem{lem}[thm]{Lemma}
\newtheorem{prop}[thm]{Proposition}
\newtheorem{defn}[thm]{Definition}
\theoremstyle{remark}
\newtheorem{rem}[thm]{Remark}
\numberwithin{equation}{section} \makeatletter
\def\@cite#1#2{#1\if@tempswa , #2\fi}
\def\@biblabel#1{$^{\hbox{\scriptsize{#1}}}$}
\makeatother
\newcommand{\R}{\mathbb R}

\newcommand{\norm}[1]{\left\Vert#1\right\Vert}

\newcommand{\set}[1]{\left\{#1\right\}}

\newcommand{\A}{\mathcal{A}}

\newcommand{\N}{\mathbb N}
\newcommand{\Z}{\mathbb Z}

\newcommand{\ep}{\epsilon}
\newcommand{\pr}{^{\prime}}

\newcommand{\beq}{\begin{equation}}
\newcommand{\eeq}{\end{equation}}
\newcommand{\ben}{\begin{enumerate}}
\newcommand{\een}{\end{enumerate}}

\newcommand{\C}{\mathbb C}

\newcommand{\eqdef}{\ensuremath{\stackrel{\mbox{\upshape \tiny
def}}{=}}} \font\sbi=cmmib10 \font\bi=cmmib10 scaled \magstep1

\begin{document}

\title[]{Solution of a linearized model of\\
Heisenberg's fundamental equation I}

\author{S. Nagamachi \and  E. Br\"uning}

\address[S. Nagamachi]{
Department of Applied Physics and Mathematics, Faculty of
Engineering, The University of Tokushima\\ Tokushima 770-8506,
Japan} \email{shigeaki@pm.tokushima-u.ac.jp}

\address[E. Br\"uning]{School of Mathematical Sciences, University
of KwaZulu-Natal, Private Bag X54001,
Durban 4000, South Africa} \email{bruninge@ukzn.ac.za}


\begin{abstract}
Heisenberg's unsolved fundamental equation of the universe [\cite{He25,He66}] has a coupling
constant $l$ which has the dimension of  length [L].  We
consider a linearized version of Heisenberg's fundamental equation which also
contains a coupling constant $l$ with the dimension of a length and we solve this
equation in the framework of a relativistic quantum field theory with a fundamental length
$\ell$ in the sense of our recently developed theory [\cite{BN04}] and show that
then one has $\ell = l/(\sqrt{2}\pi )$. This is done in two parts.
In this first part we use path integral methods (and nonstandard analysis) to calculate
all Schwinger- and all Wightman- functions of this model, as tempered
ultrahyperfunctions and verify some of the defining conditions of a relativistic quantum field
theory with a fundamental length, FLQFT for short. As an important intermediate step the convergence
of the lattice approximations for a free scalar field and for a Dirac field is shown.

The second part completes the verification of the defining conditions of FLQFT
and offers an alternative way to calculate all Wightman functions of the theory.
\end{abstract}

\maketitle

\tableofcontents
\section{Introduction}
\subsection{Heisenberg's fundamental equation}
The basic relativistic equation of quantum mechanics called Dirac
equation \beq \label{D-eq}
 i \frac{\hbar}{c}\gamma_{\mu} \frac{\partial}{\partial x_{\mu}}
\psi(x) - m\psi(x)=0, \  x_{0} = ct, x_{1} = x, x_{2} = y, x_{3} = z
\eeq contains a constants $c$ (velocity of light) which is the
fundamental constant in  relativity theory, and Planck's constant $h
= 2\pi \hbar $  which is the fundamental constant in quantum
mechanics.  The dimension of $c$ is [LT$^{-1}$] and that of $h$ is
[ML$^{2}$T$^{-1}$].  W. Heisenberg thought that a fundamental
equation of Physics must also contain a constant $l$ with the
dimension of length [L].  If such a constant $l$ is introduced, then
the dimensions of any other quantity can be  expressed in terms of
combinations of the basic constants $c$, $h$ and $l$, e.g., time [T]
= [L]/[LT$^{-1}$], or mass as [M] =
[ML$^{2}$T$^{-1}$]/([LT$^{-1}$][L]).

In 1958, Heisenberg and Pauli introduced the equation
\beq \label{HP-eq}
\frac{\hbar}{c} \gamma_{\mu} \frac{\partial}{\partial x_{\mu}}\psi(x)
\pm  l^{2} \gamma_{\mu}\gamma_{5}\psi(x)\bar{\psi}(x)\gamma^{\mu}
\gamma_{5}\psi(x)= 0,
\eeq
which was later called the {\it equation of the universe} and studied in
[\cite{DHMSY59, He66}].  The constant $l$ has the dimension [L] and is
called the fundamental length of the theory.

Unfortunately, nobody has been able to solve this equation. At
present, even in the more advanced framework of ultra-hyperfunction
quantum field theory, we do not see how this equation could be
solved. Accordingly we study a linearized version of this equations
which inherits the important property of a fundamental length $l$
and which first has been studied by Okubo [\cite{Ok61}]. This
linearized version is solvable in the sense of classical field
theory, i.e., the classical fields $\phi(x)$ and $\psi(x)=\psi\pr(x)e^{il^2 \phi(x)^2}$
solve this system when $\phi$ is a solution of the Klein-Gordon equation and
$\psi\pr$ a free Dirac field of mass $\tilde{m}$. We write it in the form  \beq \label{Oku-eq} \left\{
\begin{array}{l}
   \displaystyle \Box \phi(x)+\left(\frac{cm}{\hbar}\right)^{2}\phi(x)=0 \\
   \displaystyle \left(i\frac{\hbar}{c}\gamma^{\mu}\frac{\partial}
   {\partial x^{\mu}}-\tilde{m}\right)\psi(x)+2\gamma^{\mu}l^{2}\psi(x)\phi(x)
   \frac{\partial \phi(x)}{\partial x^{\mu}}=0
\end{array}
\right. \eeq and propose to solve the quantized version of these equations in the framework of a relativistic quantum field theory with a fundamental length as proposed recently by the authors [\cite{BN04}] by  constructing the Schwinger functions of the fields $\phi (x)$ and $\psi(x)$.
And we do so by invoking nonstandard analysis and path integral methods.
Thus we calculate the Schwinger functions by means of  path integrals on
the $*$-finite lattice with an infinitesimal spacing.  As a result, the Wightman functions (i.e., the Wick rotated Schwinger functions) of the field $\psi (x)$ are not
 tempered distributions, but an tempered ultra-hyperfunction.

In the following we will work with the natural units $c = \hbar
= 1$. Then the system of equations (\ref{Oku-eq}) reads
\begin{align} \label{Oku-eq1}
(\Box + m^{2})\phi(x)&=0\\
\label{Oku-eq2} (i\gamma^{\mu}\partial_{\mu} -
\tilde{m})\psi&=2l^{2}\gamma^{\mu}\psi(x) \phi(x)\partial_{\mu}\phi
(x).
\end{align}
and they are the field equations of the following Lagrangian density:
\begin{align}
 L(x)& = L_{Ff}(x) + L_{Fb}(x) + L_{I}(x),\label{Lagrangian}\\
 L_{Ff}(x)&=\bar{\psi}(x)(i\gamma_{\mu}\partial^{\mu}-\tilde{m})\psi(x), \label{FfLagrange} \\
  L_{Fb}(x)&=\frac{1}{2}\{(\partial^{\mu}\phi(x))^{2} - m^{2}\phi(x)^{2}\} ,\label{FbLagrange}\\
 L_{I}(x)&=2l^{2}(\bar{\psi}(x)\gamma_{\mu}\psi(x))\phi(x)\partial^{\mu}\phi(x). \label{intLagr}
\end{align}

\subsection{Relativistic quantum field theory with fundamental length (FLQFT)}
As indicated above we are going to show that the system (\ref{Oku-eq1}) - (\ref{Oku-eq2}) can be solved in the framework of a relativistic quantum field theory with a fundamental length (FLQFT) as developed in [\cite{BN04}]. This theory is essentially a relativistic quantum field theory in the sense of G{\aa}rding and Wightman [\cite{SW64}] in terms of operator-valued tempered ultra-hyperfunctions instead of operator-valued tempered Schwartz distributions. The localization properties (in co-ordinate space) of tempered ultra-hyperfunctions (for a technical explanation we have to refer to [\cite{BN04}]) are very different from those of Fourier hyperfunctions and (tempered) Schwartz distributions. Tempered ultra-hyperfunctions  distinguish events in space-time only when their distance from each other is greater than a certain length $\ell$ (A heuristic explanation of this
property is given in [\cite{BN04}]). In contrast to this, Fourier hyperfunctions and Schwartz distributions form a sheaf over space-time and thus exhibit essentially classical localization properties. On the other side the Fourier transforms of tempered ultra-hyperfunctions have essentially classical localization properties in energy-momentum space. Accordingly, compared with relativistic quantum field theory in the sense of G{\aa}rding and Wightman (abbreviated as QFT), it is the locality condition (condition of local commutativity) which needs a new formulation in FLQFT. Based on the notion of carrier of analytical functionals we proposed and used in [\cite{BN04}] the notion of {\it extended causality} or {\it extended local commutativity}.

With this notion of extended local commutativity a full set of defining conditions for a relativistic quantum field theory with a fundamental length has been given and such theories have been characterized in terms of a corresponding full set of conditions on their sequences of vacuum expectation values ($n$-point or Wightman functionals). In addition an explicit model for such a theory is constructed in [\cite{BN04}]. This model is the (Wick) exponential of the square of a free massive field $\phi$, i.e., the field
$$\rho(x)=:e^{g \phi(x)^2}:=\sum_{n=0}^{\infty} \frac{g^n}{n!}:\phi(x)^{2n}:.$$
The two-point functional of this field is formally
$$(\Omega,\rho(x)\rho(y) \Omega) = [1-4 g^2 D_m^{(-)}(x-y)^2]^{-1/2}$$
where $D_m^{(-)}(x-y)$ is the two-point functional of the field $\phi$, and the fundamental length of this model is
$$\ell = \frac{\sqrt{g}}{\pi \sqrt{2}}.$$

The major achievements of QFT are the proof of the PCT theorem, the relation between spin and statistics and the existence of a scattering matrix. In FLQFT the PCT and the spin-statistics theorems and the existence of a scattering matrix have been proven too.


\subsection{Motivation for FLQFT}
Very briefly we recall our motivation for our version of a relativistic quantum field theory with a fundamental length.

The first question one has to answer is on which level of the theory the fundamental length should be realized.

The established answer to this question is that the fundamental length should be realized on the level of the geometry of the underlying realization of space-time
and accordingly the `standard' approach to a (quantum) field theory with a fundamental length is to invoke non-commutative geometry [\cite{Co86,Wi86,Co94}].

We think that it is important to keep as many of the established physical concepts and results based on the traditional realization of space-time as possible and accordingly have proposed in [\cite{BN04}] to realize the fundamental length on the level of the primary dynamical quantities of the theory, namely the fields. In this way we can rely directly on the established physical principles (of field theory, relativistic covariance, physical energy-momentum spectrum, quantum physics). As pointed out above then the only change necessary is that of the realization of the locality principle of standard QFT (when the type of generalized functions to be used in this theory is set to be tempered ultra-hyperfunctions).
In this way we arrive at a relativistic quantum field theory in which the fundamental length is realized through special localization properties of the fields and in which the major achievements of standard QFT are still valid.

In the second part where we actually prove these localization properties for our solution we give a brief technical explanation of the localization properties of tempered ultra-hyperfunctions (see subsection 1.2).

\section{Path integral quantization}
As announced we quantize this model by path integral methods.  Formally, the time-ordered two
point function is calculated as (see [\cite{Da93}])
$$\int\bar{\psi}_{\alpha}(x_{1})\psi_{\beta}(x_{2})\exp i\left\{
\int_{\R^{4}}L_{I}(x) dx\right\}d{\mathcal D}(\psi,\bar{\psi})
d{\mathcal G}(\phi)$$
$$\times\left\{\int\exp i\left\{\int_{\R^{4}}L_{I}(x)dx\right\}
d{\mathcal D}(\psi,\bar{\psi})d{\mathcal G}(\phi)\right\}^{-1},$$
$$d{\mathcal G}(\phi)=\exp i\left\{\int_{\R^{4}}L_{Fb}(x)dx\right\}
\prod_{x\in\R^{4}}d\phi(x)$$
$$d{\mathcal D}(\psi,\bar{\psi})=\exp i\left\{\int_{\R^{4}}L_{Ff}(x)
 dx\right\}\prod_{x\in\R^{4}}\prod_{\alpha 1}^{4}\psi_{\alpha}(x)
 \bar{\psi}_{\alpha}(x).$$
All these integrals have a rigorous meaning if the continuum
space-time is replaced by a lattice. We will control the transition
from the lattice to the continuum limit by methods from non-standard
analysis.

For positive integers  $M, N$ define $L = MN$ and $\Delta  = \sqrt{\pi }/M$.
Then the lattice $\Gamma=\Gamma(M,N)$ is
 $$ \Gamma =\{ t = j\Delta ; j \in  \Z, -L < j \leq  L\} .$$
The lattice version of the differential operator
 $-\triangle  + m^{2}$ on $\R^{\Gamma ^{4}} = \R^{4\cdot 2L}$
is the following difference operator on  $\Gamma ^{4}$:
\begin{multline*}
  -\triangle + m^{2}: \R^{\Gamma^{4}} \ni  \Phi(x) \rightarrow \\ -
  \sum_{\mu =0}^{3}\frac{\Phi(x + e_{\mu}) + \Phi (x - e_{\mu}) - 2\Phi(x)}{\Delta ^{2}}+m^{2}\Phi (x) \in  \R^{\Gamma ^{4}},$$
\end{multline*}
where $e_{\mu }$ is the vector of length $\Delta $ parallel to the
$\mu $-th coordinate axis. Let $dG(\Phi )$ be a Gaussian measure on
$\R^{4\cdot 2L}$ defined by \beq \label{G-meas-bos}
dG(\Phi)=Ce^{\left\{\frac{1}{2} \sum_{y \in \Gamma^{4}}\Phi(y)
\left[\sum_{\mu
=0}^{3}\frac{\Phi(y+e_{\mu})+\Phi(y-e_{\mu})-2\Phi(y)}
{\Delta^{2}}\right. \right.
 \left. \left.-m^{2}\Phi(y)\right] \Delta^{4}\right\}} \prod_{y\in
\Gamma^{4}}d\Phi(y),
\eeq
where $C$ is the normalization constant such that $\displaystyle \int dG(\Phi ) = 1$.
 Note that the exponent of this measure is the (Euclideanized; $x^{0} \rightarrow  -iy^{0}$, $\mbox{\bi x} \rightarrow  \mbox{\bi y}$)
discretization of the Lagrangian $\displaystyle \int L_{Fb}(x) dx$.
For later use we recall the following well-known formulae for
Gaussian integrals on $\R^{4\cdot 2L}$ (see [\cite{GV64}]).
$$(2\pi)^{-n/2}\sqrt{\det \Lambda} \int e^{i(y,x)} \exp \left[-\frac{1}{2}(x,
\Lambda x)\right]dx=\exp\left[-\frac{1}{2}(y, \Lambda^{-1}y)\right]  \eqno{(A)}$$
$$(2\pi)^{-n/2} \sqrt{\det \Lambda} \int(x,Ax) \exp \left[-\frac{1}{2}(x,\Lambda x)
\right]dx={\rm Tr}(A\Lambda^{-1})\eqno{(B)}$$ where ${\rm Re
\,}\Lambda $ is strictly positive-definite and $A$ is an arbitrary
matrix. Note that the path integral on the finite lattice is the
usual integral.

Using (B), we calculate the covariance of the measure $dG(\Phi )$.
We define a function $\delta (y)$ on $\Gamma ^{4}$ by $\delta (y) =
\Delta ^{-4}$ if $y = 0$ otherwise $\delta (y) = 0$, i.e., $\delta
(y) = \Delta ^{-4}\delta _{0,y}$.  Then $(-\triangle _{x} +
m^{2})\delta (x - y)$ is the kernel function of the operator
$-\triangle +m^{2}$ and $(- \triangle _{x} + m^{2})\delta (x -
y)\Delta ^{4}\Delta ^{4}$ corresponds to the matrix $\Lambda $ of
the formulae (A) and (B) since the summation $\displaystyle \sum
_{y\in \Gamma ^{4}}$ is always accompanied by $\Delta ^{4}$.  The
inverse matrix $\Lambda ^{-1}$ corresponds $(-\triangle _{x} +
m^{2})^{-1}\delta (x - y)$ (note that there are no additional
$\Delta $).  In fact, $(-\epsilon \triangle _{x} + m^{2})\delta (x -
y)\Delta ^{4}\Delta ^{4}$ for $\epsilon  = 0$ is $m^{2} \delta (x-y)
\Delta ^{4}\Delta ^{4}$ = $m^{2} \delta _{0, x-y} \Delta ^{-4}
\Delta ^{4}\Delta ^{4}$ and its inverse is $m^{-2} \delta _{0, x-y}
\Delta ^{-4} = m^{-2} \delta (x-y)$.  Now we can calculate the
covariance of $dG(\Phi )$.
$$\int \Phi(y_{1})\Phi(y_{2}) dG(\Phi)=\Lambda^{-1}_{y_{1},y_{2}}=
(-\triangle + m^{2})^{-1}(y_{1}, y_{2}) = {\mathcal S}_{m}(y_{1} - y_{2}).$$
Using the lattice Fourier transformation, ${\mathcal S}_{m}(y_{1}
- y_{2})$ is representable as follows:
\beq \label{G-cov}
 {\mathcal S}_{m}(y_{1}-y_{2})=(2\pi)^{-4} \sum_{p\in \tilde{\Gamma}^{4}}
 e^{ip(y_{1}-y_{2})} \left[ \sum_{\mu =0}^{3}(2-2 \cos p_{\mu}\Delta)/
 \Delta^{2}+m^{2}\right]^{-1} \eta^{4},
 \eeq
 where the dual
lattice $\tilde{\Gamma}$ is given by
$$\tilde{\Gamma} =\{s=j\eta; j \in \Z, -L < j \leq L\}
,\qquad \eta =\sqrt{\pi}/N.$$ It converges, as $M, N \rightarrow
\infty $, (see the following section) to the two point Schwinger
function \beq \label{2ptSfb}
S_{m}(y_{1}-y_{2})=(2\pi)^{-4}\int_{\R^{4}}e^{ip(y_{1}-y_{2})}
\left[p^{2}+m^{2}\right]^{-1} d^{4}p. \eeq
 of a neutral scalar field of mass $m$.\\[2mm]

In order to deal with the fermion field $\Psi$ in the system
(\ref{Oku-eq1}) - (\ref{Oku-eq2}) we need to do integration over
Grassmann algebras, see [\cite{Be66}]. Accordingly we define a
measure $dD(\Psi ^{1}, \Psi ^{2})$ on the Grassmann algebra
generated by (see [\cite{Be66}]) $\{\Psi^{1}_{\alpha}(y),
\Psi^{2}_{\alpha}(y); \alpha = 1, \ldots , 4, \  y\in \Gamma ^{4}\}
$:

\begin{align} \label{G-meas-ferm}
 dD(\Psi^{1},\Psi ^{2})=&C^{\prime} e^{-\left\{\sum_{y\in \Gamma ^{4}}
 \Psi^{2T}(y)\left[\sum_{\mu=0}^{3}\gamma^{E}_{\mu}\nabla_{\mu}+\tilde{m}\right]
  \Psi^{1}(y)\Delta^{4}\right\}}   \nonumber \\
  &\prod_{y\in \Gamma^{4}}\prod_{\alpha =1}^{4}d\Psi^{1}_{\alpha}(y)
 d\Psi^{2}_{\alpha}(y),
\end{align}
 where $C^{\prime }$ is another normalization constant, and
$$ \Psi^{1}=(\Psi^{1}_{1},\ldots, \Psi^{1}_{4})^{T}, \  \Psi^{2}=(\Psi^{2}_{1},
 \ldots , \Psi^{2}_{4})^{T}.$$
 The matrices $\gamma^{E}_{\mu}$ are related to the Pauli matrices $\sigma_{j}$
 by ($j=1,2,3$)
$$ \gamma^{E}_{0}=\gamma_{0}=\left(\begin{array}{ccccc}\sigma_{0}&0\\{} 0&
-\sigma_{0}\end{array}\right), \ \gamma^{E}_{j}=-i\gamma_{j} =
\left(\begin{array}{ccc} 0&-i\sigma _{j}\\{} i\sigma _{j}&0\end{array} \right),
 $$
$$\sigma_{0}=\left(\begin{array}{ccccc} 1&0\\{} 0&1\end{array} \right) , \
 \sigma _{1} = \left( \begin{array}{ccc} 0&1\\{} 1&0\end{array} \right),\;
 \sigma_{2}=\left(\begin{array}{ccccc} 0&-i\\{} i&0\end{array} \right),\;
 \sigma_{3}=\left(\begin{array}{ccc} 1&0\\{} 0&-1\end{array} \right),$$
and the operators $\nabla_{\mu}$ as discrete versions of the
corresponding partial derivatives $\partial_{\mu}$ are defined as
follows:
$$\nabla_{\mu}\Psi_{k}=\left\{\begin{array}{cc} \nabla_{\mu}^{+}\Psi_{k}(y)=
     (\Psi_{k}(y+e_{\mu})-\Psi_{k}(y))/\Delta & {\rm if \,} \ k = 1, 2, \\
     \nabla_{\mu}^{-}\Psi_{k}(y)=(\Psi_{k}(y)-\Psi_{k}(y-e_{\mu}))/\Delta &
     {\rm if \, } \  k = 3, 4;
\end{array} \right.
  $$
namely,
$$\nabla_{\mu}=P_{+}\nabla^{+}_{\mu} + P_{-}\nabla^{-}_{\mu},
\qquad P_{\pm}= (1 \pm \gamma^{E}_{0})/2.
$$
The idea to replace the partial derivatives in the
continuum case by the forward-, respectively backward
difference on the lattice as described above, has
originally been developed in [\cite{NM86a}].

\begin{rem} \label{doubling-problem}
It is well known that the free fermion theory on the lattice
$\Gamma^{4}$ defined by the action \beq \label{fermionAction}
 \sum_{x\in \Gamma^{4}}\Psi^{2}(x)\left(\sum^{3}_{\mu =0}\gamma^{E}_{\mu}
[\Psi^{1}(x + e_{\mu})-\Psi^{1}(x - e_{\mu})]/2\Delta + m \Psi^{1}(x)\right)
 \Delta^{4},
\eeq
  suffers from the doubling problem.  Wilson
[\cite{Wi75}] has overcome this problem by adding the term
$$      - \sum _{x\in \Gamma ^{4}}\Psi ^{2}(x)\left( \sum ^{ 3}_{\mu =0}[\Psi ^{1}(x + e_{\mu }) + \Psi ^{1}(x - e_{\mu }) - 2\Psi ^{1}(x)]/2\Delta \right)  \Delta ^{4}$$
to (\ref{fermionAction})

 It is also known that the doubling
problem is due to the replacement of the partial derivative
$\partial_{\mu }$ by the central difference
$(\Psi(x+e_{\mu})-\Psi(x-e_{\mu }))/2\Delta $.  If we replace
$\partial_{\mu}$  by the forward difference $\nabla^{+}_{\mu}$
respectively the backward difference $\nabla^-_{\mu}$ as we have
suggested above, we have no doubling problems. Concretely, this is
implemented in the Fermion Lagangian density (\ref{fermionAction})
by choosing the forward difference for the components $\Psi_{1},
\Psi_{2}$ of the Fermi field while the backward difference is used
for the remaining components $\Psi_{3}, \Psi_{4}$.

Kogut and Susskind [\cite{KS75}] replaced the derivative of the
space variable by  half of the central difference, i.e., by $(\Psi
(x+e_{\mu }/2) - \Psi (x-e_{\mu }/2))/\Delta $.  Then the doubling
problem disappears but we must introduce the even lattice $\Gamma
_{e}$ and the odd lattice $\Gamma _{o}$ and assign the subset
$(\Gamma _{e} \cup \Gamma _{o})\times (\Gamma _{e} \cup \Gamma
_{o})\times \Gamma _{e})$ of $(\Gamma _{e} \cup \Gamma _{o})^{3}$ to
each field component as its domain of definition.  For further
details about lattice fermion see  [\cite{MM94}].
\end{rem}

\indent In Section 4 we are going to show that the continuum limit
of the covariance (two point function of the lattice Dirac field)
\begin{align} \label{2pfermfunc}
&{\mathcal R}_{\tilde{m} ;\alpha, \beta}(y_{1}-y_{2})= \nonumber\\
&\left[ \sum_{\mu=0}^{3}\gamma^{E}_{\mu}\nabla_{\mu}+\tilde{m}
\right ] ^{-1}_{\alpha,\beta} (y_{1},
y_{2})=\int\Psi^{1}_{\alpha}(y_{1})\Psi^{2}_{\beta}(y_{2})
dD(\Psi^{1},
 \Psi ^{2})\end{align}
coincides with the Schwinger function $R_{\tilde{m} ;\alpha,\beta}$
of the free Dirac field of mass $\tilde{m} $:
$$ R_{\tilde{m};\alpha,\beta}(y)=\left\{-\sum_{\mu=0}^{3}\gamma^{E}_{\mu}\left(
\frac{\partial}{\partial y_{\mu}}\right)+\tilde{m}\right\}_{\alpha,\beta}
S_{\tilde{m}}(y)$$ where
$$ S_{\tilde{m}}(y)= (2\pi)^{-4}\int_{\R^{4}} e^{ipy}[p^{2}+\tilde{m}^{2}]
^{-1} d^{4} p $$
\begin{rem} \label{convergence}
Though nobody seems to doubt  the convergence of the lattice approximations
 ${\mathcal S}_{m}(y_{1}-y_{2})$ respectively ${\mathcal R}_{\tilde{m};\alpha,
  \beta }(y_{1} - y_{2})$ to their standard continuum forms $ S_{m}(y_{1} - y_{2})$
  respectively $R_{\tilde{m} ;\alpha , \beta }(y_{1} - y_{2})$
  we could not find a proof. Maybe these convergence proofs are considered
  to be too tedious,  especially in the case of Fermions due to the doubling
problem.

In  Section 3, we prove this convergence by using nonstandard
analysis [\cite{Ro66, Da77}], that is we show that for any
infinitely large $M, N \in  {}^{*} \N$, the standard part of
${\mathcal S} _{m}(y_{1} - y_{2})$ is $S_{m}(y_{1} - y_{2})$.
\begin{rem} \label{nonstandard}
Sometimes, the proof by nonstandard analysis is simpler and clearer
than the standard proof. For example, in order to prove $\lim _{n
\rightarrow  \infty } f(n) = \infty $, for a function $f: \N
\rightarrow  \N$, we must show: $\forall\, M\; \exists \,N\;
\forall\, n \;(n \geq  N\; \Rightarrow  f(n) \geq  M)$.  But in
nonstandard analysis, we can use the formula $I(x)$: $x$ is an
infinitely large number, and we have only to show $\forall\, n
\;(I(n) \,\Rightarrow \, I(f(n)))$. The number of quantifiers is
reduced in this nonstandard proof, and thus it is simpler and
clearer (in technical terms: the syntactic complexity of the formula
is reduced from a $\Pi _{3}$ formula to a $\Pi _{1}$ formula.  See
[\cite{Wo05}]).
\end{rem}
In Section 4, the continuum limit of ${\mathcal R}_{\tilde{m}
;\alpha,\beta}(y_{1}-y_{2})$ is shown to be $R_{\tilde{m} ;\alpha
,\beta }$. As announced our prescription for avoiding the doubling
problem works well here.

Note that in sections 3 and 4,  convergence of Schwinger functions
is meant not in the sense generalized functions but in the sense
 of functions.

 Certainly, readers can skip sections 3 and 4 if they know or accept that these lattice approximations
 converge to their expected continuum limit.
\end{rem}
Our overall strategy is as follows:
\ben
 \item Construct the Schwinger functions $$[1 - 4l^{4} {\mathcal S}_{m}(y_{1} - y_{2})^{2}]^{-1/2}
{\mathcal R}_{\tilde{m} }(y_{1} - y_{2})$$ in the nonstandard universe;
\item by taking the standard part or continuous limit, we get standard
Schwinger functions $[1 - 4l^{4} S_{m}(y_{1} - y_{2})^{2}]^{-1/2}R_{\tilde{m} }(y_{1} - y_{2})$.
Sections 3 resp. 4 treat the continuous limit of ${\mathcal S}_{m}(y_{1} - y_{2})^{2}$ resp. $
{\mathcal R}_{\tilde{m} }(y_{1} - y_{2})$.
\item by Wick rotation, we try to define Wightman functions
$$ \lim_{\epsilon  \rightarrow  +0}[1 - 4l^{4} D_{m}^{(-)}(x_{0} - i\epsilon , \mbox{\bi x})^{2}]^{-1/2} = \lim_{\epsilon  \rightarrow +0}[1 - 4l^{4} S_{m}(ix_{0}+\epsilon , \mbox{\bi x})^{2}]^{-1/2}$$
from Schwinger functions.  But unfortunately, for this to give a mathematically well defined generalized
functions $\epsilon $ cannot be too
small, actually $\epsilon $ must be greater than $\ell  = l/(\sqrt{2}\pi )$;
\item  in this way, the Wightman functions cannot be a tempered distribution but
they can be ultra-hyperfunctions which satisfy, as we will prove later,  axiom (R0), and $\ell $ is
the fundamental length according to axiom (R3).
\een

Next we describe our strategy of how to deal with the interaction in this model.
 We define the Euclideanized lattice Lagrangian density $L_{I}(y)$ which corresponds
  to the interaction Lagrangian $L_{I}(x)$ in (\ref{intLagr})
 as follows:
$$-L_{I}(y)=\Psi^{2T}(y)e^{il^{2}\Phi(y)^{2}} \sum_{\mu=0}^{3}\gamma^{E}_{\mu}$$
$$\times [P_{+}\Psi^{1}(y + e_{\mu})\{ e^{-il^{2}\Phi(y+e_{\mu})^{2}} - e^{-il^{2}
\Phi(y)^{2}}\} /\Delta ]$$
$$ + P_{-}\Psi^{1}(y - e_{\mu})\{e^{-il^{2}\Phi(y)^{2}}-e^{-il^{2}\Phi
(y-e_{\mu})^{2}}\} /\Delta .$$ If we replace the differences  in
this definition by the corresponding partial derivatives (continuous
limit) the above Lagrangian density $L_{I}(y)$ becomes the
Euclideanization ($x^{0} \rightarrow  -iy^{0}$, $\mbox{\bi x}
\rightarrow  \mbox{\bi y}$) of $iL_{I}(x)$ as given in
(\ref{intLagr}).

Now we calculate the lattice version of the Schwinger functions of
the interacting fields.  The two point Schwinger function is

\begin{align}\label{interact2pt}
&\int \Psi^{1}_{\alpha}(y_{1})\Psi^{2}_{\beta}(y_{2})\exp\left(
\sum_{y\in \Gamma^{4}}L_{I}(y) \Delta^{4}\right) dD(\Psi^{1},\Psi^{2}) dG(\Phi)
\nonumber \\
&  \times\left\{\int \exp\left(\sum_{y\in \Gamma^{4}}L_{I}(y)\Delta^{4}\right)
dD(\Psi^{1},\Psi^{2})dG(\Phi)\right\}^{-1}.
\end{align}
If we change the variables
$$ \Psi^{1}(y)=e^{il^{2}\Phi(y)^{2}}\Psi^{\prime 1}(y), \  \Psi^{2}(y)=e^{-il^{2}\Phi(y)^{2}}
\Psi^{\prime 2}(y),$$ then (\ref{interact2pt}) becomes
$$ \int e^{il^{2}\Phi(y_{1})^{2}}\Psi^{\prime 1}(y_{1})e^{-il^{2}\Phi(y_{2})^{2}}
\Psi^{\prime 2}(y_{2}) dD(\Psi^{\prime 1},\Psi^{\prime 2}) dG(\Phi )$$
$$ = \int \Psi^{\prime 1}(y_{1})\Psi^{\prime 2}(y_{2}) dD(\Psi^{\prime 1},
\Psi^{\prime 2})\int e^{il^{2}\Phi(y_{1})^{2}}e^{-il^{2}\Phi(y_{2})^{2}} dG(\Phi ).$$
As we are going to show the continuum limit of
$$\displaystyle\int\Psi^{\prime 1}(y_{1})\Psi^{\prime 2}(y_{2})dD(\Psi^{\prime 1},
\Psi^{\prime 2})$$ is the two point Schwinger function
$R_{\tilde{m};\alpha,\beta}(y_{1}-y_{2})$ of the free Dirac field.
Abbreviate  $h_{\pm}=e^{\pm i\pi /4}l$ and observe that the characteristic
function of ($h_{-}\Phi(y_{1}),h_{+}\Phi(y_{2})$) is
$$\int e^{ith_{-}\Phi(y_{1})}e^{ish_{+}\Phi(y_{2})}dG(\Phi)=\int e^{ith_{-}\Phi
 (y_{1})+ish_{+}\Phi (y_{2})} dG(\Phi)$$
$$ =\exp -\frac{1}{2}\{th_{-}{\mathcal S}_{m}(y_{1}-y_{1})th_{-} + sh_{+}
{\mathcal S}_{m}(y_{2}-y_{1})th_{-} $$
$$+ th_{-}{\mathcal S}_{m}(y_{1}-y_{2})sh_{+}+sh_{+}{\mathcal S}_{m}(y_{2}-y_{2})
sh_{+}\} $$
$$=\exp -\frac{1}{2}\{2ts l^{2}{\mathcal S}_{m}(y_{1}-y_{2})-it^{2}l^{2}{\mathcal S}
 _{m}(0) + is^{2}l^{2}{\mathcal S}_{m}(0)\} .$$
By using the relation (see formula (A))
$$ (2\pi)^{-1/2}\int e^{ith_{\pm}\Phi(y)}e^{-t^{2}/2}dt=e^{-h_{\pm}^{2}
\Phi(y)^{2}/2} = e^{\mp il^{2}\Phi(y)^{2}/2}$$
we find
$$\int e^{il^{2}\Phi (y_{1})^{2}}e^{-il^{2}\Phi (y_{2})^{2}} dG(\Phi )$$
$$= (2\pi)^{-1} \int dt ds e^{-t^{2}/2} e^{-s^{2}/2} \int e^{it\sqrt{2}h_{-}
\Phi(y_{1})}e^{i\sqrt{2}sh_{+}\Phi(y_{2})} dG(\Phi)$$
$$=(2\pi)^{-1} \int dt ds e^{-t^{2}/2} e^{-s^{2}/2} e^{
-\left\{ 2tsl^{2} {\mathcal S} _{m}(y_{1}-y_{2})-\left(it^{2}l^{2}
 {\mathcal S}_{m}(0)-is^{2}l^{2} {\mathcal S}_{m}(0)\right) \right\}} $$
$$= \left[(1-2il^{2}{\mathcal S}_{m}(0))(1+2il^{2}{\mathcal S}_{m}(0))-4l^{4}
{\mathcal S}_{m}(y_{1}-y_{2})^{2}\right]^{-1/2}$$ where we used
formula (A) of Gaussian integration for $y=0$.

 The value of the two-point Schwinger function at the origin in the lattice
approximation diverges in the continuum limit, i.e., ${\mathcal S}
_{m}(0)={\mathcal S}_{m}(0;N, M) \rightarrow  \infty $ as $N, M
\rightarrow  \infty $.  In fact,
$$ {\mathcal S}_{m}(0)=(2\pi)^{-4} \sum_{p\in \tilde{\Gamma}^{4}}\left[
\sum_{\mu=0}^{3}(2-2\cos p_{\mu}\Delta)/\Delta^{2}+m^{2}\right]^{-1}
(\sqrt{\pi}/N)^{4}$$
$$\geq(2\pi)^{-4}\sum_{p\in \tilde{\Gamma}^{4}}\left[ \sum_{\mu =0}^{3}
4\vert p\vert^{2}/\pi^{2}+m^{2}\right]^{-1}(\sqrt{\pi}/N)^{4}$$
$$\rightarrow(2\pi)^{-4} \int_{\vert p_{\mu}\vert \leq \sqrt{\pi}M}\left[
\sum_{\mu=0}^{3} 4\vert p\vert^{2}/\pi^{2}+m^{2}\right]^{-1} d^{4}p\
(N \rightarrow  \infty )$$
$\rightarrow  \infty $ as ($M \rightarrow  \infty $).

As usual we eliminate this divergence by
interpreting the above products respectively power series in the
sense of Wick products. A compact form to define Wick products is as
follows (see [\cite{GJ81}]):
$$ :e^{ith_{\pm}\Phi(y)}:=\sum _{n=0}^{\infty}[:(ith_{\pm}\Phi(y))^{n}:/n!]=e^{\mp it^{2}l^{2}
{\mathcal S}_{m}(0)} e^{ith_{\pm }\Phi (y)}.$$ Then we have
$$\int :e^{ith_{-}\Phi(y_{1})}: \, :e^{ish_{+}\Phi(y_{2})}: dG(\Phi)=\exp -\left\{2tsl^{2}
{\mathcal S}_{m}(y_{1} - y_{2}) \right\} $$
and
$$ \int :e^{il^{2}\Phi(y_{1})^{2}}: \,  :e^{-il^{2}\Phi(y_{2})^{2}}: dG(\Phi )$$
$$ = (2\pi )^{-1} \int dt ds e^{-t^{2}/2} e^{-s^{2}/2} \int :e^{i\sqrt{2}th_{-}\Phi(y_{1})}: \,
:e^{i\sqrt{2}sh_{+}\Phi (y_{2})}: dG(\Phi)$$
$$ =(2\pi)^{-1} \int dt ds e^{-t^{2}/2} e^{-s^{2}/2} \exp -\left\{2tsl^{2}{\mathcal S}_{m}
(y_{1} -y_{2}) \right\} $$
$$ =\left[1 - 4l^{4} {\mathcal S}_{m}(y_{1} - y_{2})^{2}\right]^{-1/2}.$$
Thus the two point Schwinger function of the field $\psi $ in
lattice approximation is
$$ \left[1 - 4l^{4} {\mathcal S} _{m}(y_{1} - y_{2})^{2}\right]^{-1/2} {\mathcal R}_{\tilde{m} ;
\alpha , \beta }(y_{1} - y_{2}),$$
and its continuous limit (Section 4) is
$$ \left[1 - 4l^{4} S_{m}(y_{1}- y_{2})^{2}\right]^{-1/2} R_{\tilde{m} ;\alpha,\beta}(y_{1} - y_{2}).$$
\medskip

In order to construct the complete theory the system of Schwinger
respectively Wightman functions of all orders $n \in \N$ has to be
constructed. We show here how all $n$-point functions of the
interacting fields $\phi$ and $\psi$ can be calculated in the
lattice approximation.
 For the $n$-point Schwinger functions of the fields $:e^{it_{j}h_{r_{j}}\phi (x)}:$ we find
$$\int \prod_{j=1}^{n} :e^{it_{j}h_{r_{j}}\Phi(y_{j})}: dG(\Phi)
 = \prod_{j<k}e^{- t_{j}t_{k}h_{r_{j}}h_{r_{k}} {\mathcal S}_{m}(y_{j} - y_{k})} ,$$
where $r_{j} = +$ or $r_{j} = -$, and
$$ \int \prod_{j=1}^{n} :e^{ -(-1)^{r_{j}}il^{2} \Phi (y_{j})^{2}}:
 dG(\Phi )$$
$$ =\int \prod_{j=1}^{n} dt_{j} \int \prod_{j=1}^{n} :e^{i\sqrt{2}t_{j}h_{r_{j}}\Phi(y_{j})}:
 e^{-t_{j}^{2}/2} dG(\Phi)$$
$$= \int \prod_{j=1}^{n} dt_{j}e^{-t_{j}^{2}/2} \prod_{j<k}
e^{-2\{ t_{j}t_{k}h_{r_{j}}h_{r_{k}} {\mathcal S}_{m}(y_{j} -
y_{k})} = (\det C)^{-1/2},$$ where the matrix $C= (c_{j,k})$ is
given by \beq \label{abbrev1}
 c_{j,j} = 1, \  c_{j,k} = c_{k,j} = 2h_{r_{j}}h_{r_{k}}l^{2}
 {\mathcal S}_{m}(y_{j} - y_{k})\;j < k.
\eeq where again formula (A) has been used.

 Similarly, introduce the matrix $A=(a_{j,k})$ by
 \begin{align}  \label{abbrev2}
 a_{j,j} = 1, \;  a_{j,k} = a_{k,j} &= 2h_{r_{j}}h_{r_{k}}l^{2}
 D_{m}^{(-)}(x_{j} - x_{k}),\;j < k,\\
  D_{m}^{(-)}(x_{0}, \mbox{\bi x}) &= S_{m}(ix_{0}, \mbox{\bi x}).
  \end{align}
Then  the $n$-point Wightman function
$$      \langle 0\vert  \rho ^{(1)}(x_{1}) \cdots \rho ^{(n)}(x_{n}) \vert 0\rangle $$
of the field \beq \label{field-rho}
 \rho^{(j)}(x_{j}) = :e^{-(-1)^{r_{j}}il^{2}\phi(x_{j})^{2}}:
 \eeq
is the Wick rotation of the Schwinger function (\ref{abbrev1}), i.e.,
 $$(\det A)^{-1/2}.$$ This field has been studied in some detail
in [\cite{BN03}], see also [\cite{NB03}].\\

Next, let $\psi _{0}(x)$ be the free Dirac field of mass $\tilde{m} $
and introduce the field components
$\psi ^{1}(x) = \psi _{0}(x), \ \psi ^{2}(x) = \bar{\psi } _{0}(x)$.
Denote the Wightman function of the free Dirac field $\psi _{0}(x)$ by
$${\mathcal W}^{r}_{0,\alpha}(x_{1}, \ldots , x_{n})=(\Omega,
\psi^{r_{1}}_{\alpha_{1}}(x_{1}) \cdots \psi^{r_{n}}_{\alpha_{n}}(x_{n})
\Omega )$$
  and let
$S^{r}_{0,\alpha}(y_{1}, \ldots , y_{n})$ be its Schwinger function, where
$r = (r_{1}, \ldots , r_{n})$, $\alpha  = (\alpha_{1}, \ldots, \alpha_{n})$.
Then the $n$-point Schwinger function of $\psi (y)$ is
\beq\label{nPoint-Schw}
 (\det C)^{-1/2} S^{r}_{0, \alpha}(y_{1}, \ldots , y_{n}),
 \eeq
where ${\mathcal S}_{m}(y_{j} - y_{k})$ of (\ref{G-cov}) is replaced
by its continuous limit $S_{m}(y_{j} - y_{k})$ (\ref{2ptSfb}) and
where the matric $C$ is defined in (\ref{abbrev1}).

Similarly, introduce the matrix $A= (a_{j,k})$ as in (\ref{abbrev2}).
Then the $n$-point
Wightman function ${\mathcal W} ^{r}_{\alpha }(x_{1}, \ldots , x_{n})$
of the field $\psi(x)$ is the Wick rotation of the Schwinger function (\ref{nPoint-Schw}), i.e.,
\begin{equation} \label{nPoint-Wigh}
 {\mathcal W}^{r}_{\alpha}(x_{1}, \ldots , x_{n}) =
 (\det A)^{-1/2} {\mathcal W}^{r}_{0,\alpha}(x_{1}, \ldots , x_{n}).
\end{equation}

In Section 5, we show that the Wightman functions of $\psi (x)$
are not tempered distributions but tempered ultrahyperfunctions
which are studied in [\cite{Ha61, Mo75, BN04}].  The axiom (R0) of
reference [\cite{BN04}], modified for the case of Dirac fields is
verified.

In part II of our investigations of this linearized model of Heisenberg's equation [\cite{BN07}], it is shown that the present model satisfies all the axioms of relativistic quantum field theory
with a fundamental length.

\section{Convergence of the lattice approximation of the two point functions for free scalar
fields}
For positive integers $M, N$  put $L = MN$ and let $\Gamma $ be the
1-dimensional lattice
$$\Gamma=\{x = j\Delta; j \in \Z, -L < j \leq L\},$$
with spacing $\Delta=\sqrt{\pi}/M$. Its dual lattice
$$\tilde{\Gamma}=\{p=j\eta; j \in \Z, -L < j \leq L\}$$
then has the spacing $\eta=\sqrt{\pi}/N$.
Let $e_{\mu}$ be the vector parallel to the $\mu$-th coordinate axis with
length $\Delta$, and $\nabla_{\mu}^{\pm}$ the forward respectively backward
difference in direction $e_{\mu}$ defined by
$$\nabla_{\mu}^{+}\Phi(x)=\frac{\Phi(x + e_{\mu})-\Phi(x)}{\Delta},
\quad \nabla_{\mu}^{-}\Phi(x)=\frac{\Phi(x) - \Phi(x-e_{\mu
})}{\Delta }.$$ Then we have
\begin{align}\label{forbackdiff}
\nabla_{\mu}^{+}e^{ipx}&=\frac{e^{ip_{\mu}\Delta}-1}{\Delta}e^{ipx}=
i\bar{q}_{\mu}e^{ipx}\\
 \nabla_{\mu}^{-}e^{ipx}&=\frac{1-e^{-ip_{\mu}\Delta}}
{\Delta}e^{ipx} = iq_{\mu}e^{ipx},
\end{align}
where $q_{\mu} = (1 - e^{-ip_{\mu}\Delta})/(i\Delta )$.  Note that
$$\nabla_{\mu}^{+}\nabla_{\mu}^{-}e^{ipx}=
-\vert q_{\mu}\vert^{2}e^{ipx} = - \frac{2 - 2\cos
p_{\mu}\Delta}{\Delta^{2}}e^{ipx}.$$ Accordingly we define a linear
operator $-\triangle + m^{2} =-\sum_{\mu=0}^{3}
\nabla_{\mu}^{+}\nabla_{\mu}^{-} + m^{2}$ on $\R^{\Gamma ^{4}} =
\R^{4\cdot 2L}$ (second order difference operator on the lattice
$\Gamma ^{4}$) by
\begin{multline*}
-\triangle + m^{2}:\R^{\Gamma ^{4}}\ni \Phi(x) \rightarrow \\
-\sum_{\mu=0}^{3}\frac{\Phi(x + e_{\mu}) + \Phi(x - e_{\mu}) -
2\Phi(x)} {\Delta^{2}}+m^{2}\Phi (x) \in \R^{\Gamma ^{4}}.
\end{multline*}
Using lattice Fourier transformation with periodic boundary conditions, i.e.,
$$ \tilde{\Phi}(p)=(2\pi)^{-2} \sum_{x \in \Gamma^{4}}e^{-ipx} \Phi(x) \Delta^{4},$$
$$ \Phi(x)=(2\pi )^{-2} \sum_{p \in \tilde{\Gamma}^{4}}e^{ipx} \tilde{\Phi}(p)
\eta ^{4}.$$
this operator has the following simple form in terms of the lattice Fourier transform
$\tilde{\Phi}(p)$ of $\Phi (x)$ :
$$ \tilde{\Phi}(p)\rightarrow \left(\sum_{\mu =0}^{3}\frac{- e^{ip_{\mu}\Delta }
- e^{-ip_{\mu}\Delta} + 2}{\Delta^{2}}+m^{2}\right) \tilde{\Phi}(p)$$
$$ =\left(\sum_{\mu=0}^{3} \frac{2 - 2\cos p_{\mu }\Delta}
{\Delta^{2}}+m^{2}\right) \tilde{\Phi}(p).$$
Therefore the kernel $K(x, y)$ (the matrix) of the linear operator
$-\triangle + m^{2}$ is:
$$ K(x,y)=(2\pi)^{-4} \sum_{p\in \tilde{\Gamma}^{4}} e^{ip(x - y)}
\left(\sum_{\mu =0}^{3} \frac{2 - 2\cos p_{\mu}\Delta }{\Delta^{2}} + m^{2} \right)
  \eta ^{4}.$$
and the kernel of its inverse is accordingly \beq
\label{K-inverse}
 K^{-1}(x,y)=(2\pi)^{-4} \sum_{p\in \tilde{\Gamma}^{4}}e^{ip(x - y)}
\left(\sum_{\mu =0}^{3} \frac{2 - 2\cos p_{\mu}\Delta}{\Delta^{2}}+m^{2}\right)^{-1}
 \eta ^{4}.\eeq
 Note that (\ref{K-inverse}) can be written as
\beq \label{K-inverse1} =\frac{1}{(2\pi)^3}\sum_{\mbox{\bi
p}\in \tilde{\Gamma}^{3}}e^{i\mbox{\bi p}\cdot(\mbox{\bi
x}-\mbox{\bi y})}\left(\frac{1}{2\pi}\sum_{p_0 \in
\tilde{\Gamma}}\frac{e^{ip_0(x^0 - y^0)}}{\frac{2-\cos
p_0\Delta}{\Delta^2}+ A(\mbox{\bi p})^2}\eta \right)\eta^3
\eeq where
$$A(\mbox{\bi p})^2=m^2+\sum_{\mu=1}^3
\frac{2-2\cos p_{\mu}\Delta}{\Delta^{2}}$$

Accordingly we calculate and estimate, for $x \in \Gamma $
and some $B\neq 0$ which later will be chosen  to
equal $A(\mbox{\bi p})$, the one dimensional lattice sum
\beq \label{1dlsum}
 \sum_{p \in \tilde{\Gamma}} \frac{e^{ixp}}{(2 - 2\cos
p\Delta)/\Delta^{2} + B^{2}} \eta. \eeq The result is:
\begin{prop} \label{prop:1dlsum}
 Assume $B \neq 0$ and $\vert \arg B\vert \leq  \pi /4$.  Then one has, for all
 $x \in \Gamma$,
\begin{align} \label{eq:1dlsumvalue1}
 &\sum_{p \in \tilde{\Gamma}}\frac{e^{ixp}}{(2-2\cos
p\Delta)/\Delta^{2}+ B^{2}} \eta= \frac{2\pi\Delta
z_+^{-|x|/\Delta}}{z_+ -z_-}
\\
\label{eq:1dlsumvalue2} & = \frac{2\pi(1+\Delta
B[\sqrt{4+\Delta^{2}B^{2}}/2 +\Delta B/2])^{-\vert
 x\vert /\Delta}}{B\sqrt{4+\Delta^{2}B^{2}}},\end{align}
 with $z_{\pm}$ given in (\ref{zplusminus}).

If $M, N \in {}^{*}\N$ are infinitely large numbers and, in
the case $B$ is infinitely large,  $\delta
 = \Delta B[\sqrt{4 +\Delta^{2}B^{2}}/2 + \Delta B/2]$ is
 infinitesimal, then Eq. (\ref{eq:1dlsumvalue2}) can be continued by
\beq \label{eq:1dlsumvalue1}
 =\frac{2\pi e_{*}^{-B[\sqrt{4+\Delta^{2}B^{2}}/2 +\Delta B]|x|}}
{B\sqrt{4+\Delta^{2}B^{2}}} \eeq for some $e_{*} \approx
e$, which is near $2\pi e^{-B|x|}/2B$.
\end{prop}
\begin{proof}
In order to evaluate the sum (\ref{1dlsum}) we first rewrite it as
\begin{align*}
 &\sum_{p \in \tilde{\Gamma}} \frac{e^{ixp}}{(2 - 2\cos
p\Delta)/\Delta^{2} + B^{2}} \eta
 =\sum_{p \in \tilde{\Gamma}} \frac{e^{ixp}}{(2 - e^{ip\Delta} - e^{-ip\Delta })
/\Delta ^{2} + B^{2}} \eta =\\
&\sum_{p \in \tilde{\Gamma}} \frac{e^{i(x+\Delta)p}}
{(2e^{ip\Delta}- e^{i2p \Delta} - 1)/\Delta^{2} + e^{ip\Delta}B^{2}}
\eta =\sum_{p \in \tilde{\Gamma}} \frac{\Delta^2 e^{ixp}z}{(2z - z^2
 - 1) + \Delta^2B^{2}z} \eta.\\
\end{align*}
for $z = e^{ip\Delta }$. For the decomposition into partial
fractions we determine the zeros of the denominator, as a function
of the complex variable $z$, i.e., of
$$  z^{2} - (2 + \Delta^{2}B^{2})z + 1 = 0.$$
These zeros are \beq \label{zplusminus} z=z_{\pm} =\frac{2 +
\Delta^{2}B^{2} \pm \Delta B\sqrt{4 + \Delta^{2}B^{2}}}{2};\eeq and
we can write
$$\frac{1}{2z -z^2 -1 +z\Delta^{2}B^{2}}=\frac{1}{z_+ - z_-}
\left(\frac{1}{z - z_-}-\frac{1}{z - z_+} \right).$$ Under
our assumptions for $B$ we know that ${\rm Re\,}z_{+}>1$,
$\vert z_{-}\vert < 1$ and $z_+\cdot z_-=1$. This allows us
to use a geometric series to evaluate the lattice sum.
Under these conditions we get
$$\frac{z}{z-z_-} - \frac{z}{z-z_+}=\frac{1}{1- \frac{z_-}{z}}
+\frac{z/{z_+}}{1-
\frac{z}{z_+}}=\sum_{k=0}^{\infty}\left(\frac{z_-}{z}\right)^k
+
\frac{z}{z_+}\sum_{k=0}^{\infty}\left(\frac{z}{z_+}\right)^k
$$
and accordingly the evaluation of the lattice sum is continued by
\begin{align} \label{1dlsumeval2}
\frac{\Delta^2 \eta}{z_+ -z_-}\sum_{p \in \tilde{\Gamma}}
e^{ixp}\left( \sum_{k=0}^{\infty}z_-^k e^{-ipk\Delta}
+\sum_{k=0}^{\infty} z_+^{-k-1}e^{ip(k+1)\Delta} \right)= \nonumber\\
=\frac{\Delta^2 \eta}{z_+ -z_-}\sum_{k=0}^{\infty}\left(z_-^k
\sum_{p \in \tilde{\Gamma}}e^{ixp}e^{-ipk\Delta} + z_+^{-k-1}
\sum_{p \in \tilde{\Gamma}}e^{ixp}e^{ip(k+1)\Delta}
 \right).
\end{align}
Now observe that lattice points are of the form $x=k_0 \Delta$ for
some integer $k_0$, $-L+1\leq k_0 \leq L$ while points of the dual
lattice have the form $p=j\eta$, $-L+1 \leq j \leq L$. For $m \in
\Z$ one has the following cases
$$\sum_{p \in \tilde{\Gamma}}e^{im\Delta p}=\sum_{j=-L+1}^L
e^{im\Delta j\eta}= \begin{cases} 2L &  m=0\\
e^{im\frac{\pi}{L}(-L+1)}\frac{1-e^{im\frac{\pi}{L}2L}}
{1-e^{im\frac{\pi}{L}}}=0 & m\neq 0.
\end{cases}$$
Accordingly (\ref{1dlsumeval2}) equals \begin{align*}
\frac{\Delta^2 \eta}{z_+
-z_-}\sum_{k=0}^{\infty}\left(z_-^k 2L \delta_{k_0,k} +
z_+^{-k-1} 2L \delta_{-k_0,k+1}\right)=\\= \frac{\Delta^2
\eta}{z_+ -z_-}2L\left(\theta(x)z_-^{k_0}+z_+^{k_0}
\theta(-x)\right)=\frac{\Delta^2 \eta}{z_+ -z_-}2L
z_+^{-|x|/\Delta}.\end{align*} By inserting the expression
(\ref{zplusminus}), the value (\ref{eq:1dlsumvalue2}) for
the one dimensional lattice sum follows.

Now assume that $M, N \in {}^{*}\N$ are infinitely large numbers.
Denote $u = \delta /\vert \delta \vert $, where $\delta=\Delta
B[\sqrt{4+\Delta^{2} B^{2}}/2 + \Delta B/2]$. Then we have
$$\frac{(1+\Delta B[\sqrt{4+\Delta^{2}B^{2}}/2 +\Delta B/2])^{-\vert x\vert
 /\Delta }}{B\sqrt{4+\Delta^{2}B^{2}}}=$$
$$=\frac{[(1+u\vert \delta \vert)^{1/\vert \delta \vert}]^{-\vert
\delta \vert \vert x\vert
/\Delta}}{B\sqrt{4+\Delta^{2}B^{2}}}=
\frac{e_{*}^{-B[\sqrt{4+\Delta^{2}B^{2}}/2 +\Delta
B/2]|x|}}{B \sqrt{4+ \Delta^{2}B^{2}}}$$ where we put
$$(1+u|\delta|)^{1/|\delta|}=e_*^{u},$$ i.e.,
$e_*=e^{\frac{1}{u|\delta|}\log (1+u|\delta|)}$ and
$\frac{1}{u|\delta|}\log (1+u|\delta|)\approx 1$, if
$\delta$ is infinitesimally small.
\end{proof}
\begin{rem}
Among other things our calculations for the lattice sum have
established that for $x \in \Gamma$,
$$\sum_{p \in \tilde{\Gamma}}\frac{e^{ipx}e^{ip\Delta}}{e^{ip\Delta}
-z_-}\Delta \eta = 2L\Delta \eta z_-^{x/\Delta}=2\pi z_-^{x/\Delta}.
$$
The continuum version of this result reads
$$\int_{-\sqrt{\pi}M}^{\sqrt{\pi}M} \frac{\Delta e^{i(x+\Delta)p}}
{e^{ip\Delta}-z_{-}}dp=\int_{\vert z\vert =1}
\frac{z^{x/\Delta} } {z - z_{-}}\frac{dz}{i}=2\pi
z_{-}^{x/\Delta} .$$ It is interesting to note that the
summation and the integration give precisely the same
value.
\end{rem}

\begin{prop} \label{prop3.2a}
 Let $M, N \in {}^{*}\N$ be infinitely large numbers and
$M_{0} = \sqrt{M}$.  If $\vert \mbox{\bi p}\vert \leq  M_{0}$, then for $x_0 \in \Gamma$,
\begin{align}\label{invKernel-approx1}
 &(2\pi)^{-4} \sum_{p_{0}\in \tilde{\Gamma}} e^{ipx} \left(\sum_{\mu =0}^{3}
 \frac{2 - 2\cos p_{\mu}\Delta}{\Delta^{2}}+m^{2}\right)^{-1} \eta \nonumber \\
& =(2\pi)^{-3} \frac{e^{i\mbox{\sbi p}\mbox{\sbi x}}e_{**}
(\mbox{\bi p})^{-\sqrt{\vert \mbox{\sbi q}\vert ^{2} + m^{2}}\vert x_{0}
\vert }}{2\sqrt{\vert \mbox{\sbi q}\vert ^{2} + m^{2}}},
\end{align}
and $e_{**}(\mbox{\bi p}) \approx e$.  If $\vert \mbox{\bi
p}\vert \geq  M_{0}$ then
\begin{align}\label{invKernel-approx2}
& \left|(2\pi)^{-4} \sum _{p_{0}\in \tilde{\Gamma}} e^{ipx}
\left(\sum_{\mu=0}^{3}\frac{2-2\cos p_{\mu} \Delta}{\Delta^{2}}+m^{2}
\right)^{-1}\eta \right| \nonumber \\
& \leq (2\pi)^{-3}2^{-2 M_{0}\vert x_{0}\vert /\pi}
\frac{1}{2^2 M_{0}/\pi}.
\end{align}
\end{prop}
\begin{proof}
Recall the definition of $A(\mbox{\bi p})$ in
(\ref{K-inverse1}). A basic estimate for the cosine, i.e.,
$\frac{2}{\pi^2}t^2 \leq 1- \cos t \leq \frac{1}{2}t^2$ for
$|t|\leq \pi$, yields
$$m^2 +\frac{4}{\pi^2} \mbox{\bi p}^2 \leq A(\mbox{\bi p})^2 \leq
m^2 + \mbox{\bi p}^2,$$ since $|p_{\mu}\Delta| \leq \pi$
for $p \in \tilde{\Gamma}^4$.

We prepare the application of Proposition \ref{prop:1dlsum}
with $B=A\equiv A(\mbox{\bi p})$ by checking that
   $\delta=\Delta
A[\sqrt{4 + \Delta^{2}A^{2}}/2 + \Delta A/2]$ is
infinitesimal. On the basis of the above estimate for
$A(\mbox{\bi p})^2$ this is straightforward  for $\vert
\mbox{\bi p}\vert \leq  M_{0}$. Thus
(\ref{invKernel-approx1}) follows.

Since $A\sqrt{4+\Delta^{2}A^{2}}$ and $
z_{+}=(2+\Delta^{2}A^{2}+\Delta
A\sqrt{4+\Delta^{2}A^{2}})/2$ are increasing functions of
$A \geq  0$, $z_{+}^{-\vert x_{0}\vert } /A\sqrt{4 +
\Delta ^{2}A^{2}}$ is a decreasing function of $A$ and thus
is estimated from above by its value at the minimum value
$A_0$ for $A(\mbox{\bi p})$ for $|\mbox{\bi
p}|>M_0=\sqrt{M}$. By our estimate for $A(\mbox{\bi p})^2$
it follows $A_0 \leq 2 \sqrt{M}$ and again $\Delta A_0$ and
$\delta_0=\Delta A_0[\sqrt{4 + \Delta^{2}A_0^{2}}/2 +
\Delta A_0/2]$ are infinitesimal. Hence Proposition
\ref{prop:1dlsum} applies and (\ref{invKernel-approx2})
follows from the lower bound $A_0 \geq \frac{2}{\pi}M_0$.
 \end{proof}
 Next we prepare the evaluation of the 4-dimensional lattice sum by two lemmas.
\begin{lem} \label{lem3.3}
 Let $M,N \in{}^{*}\N$ be infinitely large numbers. If $ x_{0} \in \Gamma $
is not infinitesimal, then
$$(2\pi)^{-4}\sum_{p\in \tilde{\Gamma}^{4}}e^{ipx}\left(\sum_{\mu=0}^{3}
\frac{2-2\cos p_{\mu}\Delta}{\Delta^{2}}+m^{2}\right)^{-1} \eta^{4}$$
$$ \approx(2\pi)^{-3}\sum_{\mbox{\sbi p}\in \tilde{\Gamma}^{3}}
\frac{e^{i\mbox{\sbi p} \mbox{\sbi x}}e^{-\sqrt{\vert \mbox{\sbi p}\vert^{2}+
 m^{2}}\vert x_{0}\vert}}{2\sqrt{\vert\mbox{\sbi p}\vert^{2}+m^{2}}} \eta^{3}.$$
\end{lem}
\begin{proof}
Let $M_{0}=\sqrt{M}$ and suppose that $\vert x_{0}\vert $ is not
infinitesimal. Then, by Proposition \ref{prop3.2a}
$$\left|(2\pi)^{-4}\sum_{\mbox{\sbi p}\in \tilde{\Gamma}^{3}, \vert \mbox{\sbi p}
\vert \geq M_{0}}\sum_{p_{0}\in \tilde{\Gamma}} e^{ipx}\left(\sum_{\mu=0}^{3}
\frac{2-2\cos p_{\mu}\Delta}{\Delta^{2}}+m^{2}\right)^{-1} \eta^{4}\right| $$
$$\leq (2\pi)^{-3}\sum_{\mbox{\sbi p}\in \tilde{\Gamma}^{3},
\vert \mbox{\sbi p}\vert \geq  M_{0}}\frac{2^{-\sqrt{2}M_{0}\vert
x_{0}\vert /\pi}}{2\sqrt{2}M_{0} /\pi } \eta^{3}
 \leq(2\pi)^{-3/2} M^{3}\frac{2^{-2\sqrt{M}\vert x_{0}\vert /\pi }}
{2^2\sqrt{M}/\pi} \approx 0.  $$  Since
$$ \left|(2\pi)^{-3}\frac{e_{**}(\mbox{\bi
p})^{-\sqrt{\vert\mbox{\sbi q} \vert^{2}+m^{2}}\vert
x_{0}\vert}}{2\sqrt{\vert \mbox{\sbi q}\vert^{2}+
 m^{2}}}\right|
\leq(2\pi)^{-3} \frac{2^{-2\vert \mbox{\sbi p}\vert\vert x_{0}\vert
/\pi}}{4\vert\mbox{\sbi p} \vert /\pi},$$ for any standard
$\epsilon> 0$, there exists a finite $M_{1} > 0$ such that
$$\left|(2\pi)^{-3}\sum_{\mbox{\sbi p}\in \tilde{\Gamma}^{3}, M_{1}\leq \vert
\mbox{\sbi p}\vert \leq M_{0}} \frac{e_{**}(\mbox{\bi
p})^{-\sqrt{\vert\mbox{\sbi q} \vert^{2}+m^{2}}\vert
x_{0}\vert}}{2\sqrt{\vert \mbox{\sbi q}\vert^{2}+
 m^{2}}}
 \eta^{3} \right| < \epsilon $$
if $\vert x_{0}\vert $ is not infinitesimal.  This shows that for
all $\epsilon> 0$ there exists $M_{1}$ such that
$$\left|(2\pi)^{-4} \sum_{\mbox{\sbi p}\in \tilde{\Gamma}^{3}, \vert
\mbox{\sbi p}\vert \geq M_{1}} \sum_{p_{0}\in \tilde{\Gamma}}e^{ipx}
\left(\sum_{\mu =0}^{3} \frac{2-2\cos p_{\mu}\Delta}{\Delta^{2}}+m^{2}\right)^{-1}
 \eta^{4} \right| < \epsilon .$$
We  also have
$$\forall\; \epsilon > 0\; \exists\; M_{1}\quad \left| \sum_{\mbox{\sbi p}\in
\tilde{\Gamma} ^{3},\vert \mbox{\sbi p}\vert \geq  M_{1}}
 \frac{e^{-\sqrt{\vert \mbox{\sbi p}\vert^{2}+m^{2}}\vert x_{0}\vert}}
 {\sqrt{\vert \mbox{\sbi p}\vert^{2}+m^{2}}} \eta^{3} \right|<\epsilon .$$
Let $M_{1} > 0$ be finite.  If $\vert \mbox{\bi p}\vert\leq  M_{1}$,
then one has, for some $0 < \theta_{\mu}=\theta(p_{\mu}) < 1$,
$$\vert \mbox{\bi q}\vert^{2}= \sum_{\mu=1}^3\frac{2-2\cos p_{\mu}\Delta}
{\Delta^{2}}=\sum_{\mu=1}^3 ( p_{\mu}^{2}+
 \frac{\sin \theta_{\mu} p_{\mu}\Delta}{3!} p_{\mu}^{3}\Delta) \approx \sum_{\mu=1}^3 p_{\mu}^{2} \
 $$
and
$$ = \frac{e_{**}(\mbox{\bi p})^{-\sqrt{\vert \mbox{\sbi q}\vert^{2}+m^{2}}
\vert x_{0}\vert}}{2\sqrt{\vert \mbox{\sbi q}\vert^{2}+m^{2}}}
\approx \frac{e^{-\sqrt{\vert \mbox{\sbi p}\vert^{2}+m^{2}}\vert
x_{0}\vert}} {2\sqrt{\vert \mbox{\sbi p}\vert^{2}+m^{2}}}.$$ It
follows that
$$\left|(2\pi)^{-3}\sum_{\mbox{\sbi p}\in \tilde{\Gamma}^{3},\vert \mbox{\sbi p}
\vert \leq  M_{1}} \frac{e^{i\mbox{\sbi p} \mbox{\sbi x}}e_{**}
(\mbox{\bi p})^{-\sqrt{\vert \mbox{\sbi q}\vert^{2}+m^{2}}\vert x_{0}\vert}}
{2\sqrt{\vert \mbox{\sbi q}\vert^{2}+m^{2}}}\eta ^{3} \right.$$
$$ \left. -(2\pi)^{-3} \sum_{\mbox{\sbi p}\in \tilde{\Gamma}^{3},\vert
\mbox{\sbi p}\vert \leq M_{1}}\frac{e^{i\mbox{\sbi
p}\mbox{\sbi x}}e^ {-\sqrt{\vert \mbox{\sbi
p}\vert^{2}+m^{2}}\vert x_{0}\vert}}{2\sqrt{\vert
\mbox{\sbi p}\vert^{2}+m^{2}}} \eta^{3} \right| \approx 0$$
and hence
$$ (2\pi)^{-4}\sum_{p\in \tilde{\Gamma}^{4}} e^{ipx} \left(\sum_{\mu=0}^{3}
\frac{2-2\cos p_{\mu}\Delta}{\Delta^{2}}+m^{2}\right)^{-1} \eta^{4}
$$ $$ \approx (2\pi)^{-3} \sum_{\mbox{\sbi p}\in \tilde{\Gamma}^{3}}
\frac{e^{i\mbox{\sbi p} \mbox{\sbi x}}e^{-\sqrt{\vert \mbox{\sbi
p}\vert^{2}+ m^{2}}\vert x_{0}\vert}}{2\sqrt{\vert \mbox{\sbi
p}\vert^{2}+m^{2}}} \eta^{3}.$$
\end{proof}
\begin{lem} \label{lem3.4}
Assume that $M,N \in ^*\N$ are infinitely large numbers. If
$\mbox{\bi x}$ is finite and $|x_0|$ not infinitesimal,
then the following lattice sum is infinitesimally close to
the expected integral:
$$(2\pi)^{-3}\sum_{\mbox{\sbi p}\in \tilde{\Gamma}^{3}}
\frac{e^{i\mbox{\sbi p}\mbox{\sbi x}}e^{-\sqrt{\vert \mbox{\sbi
p}\vert^{2}+m^{2}}\vert x_{0}\vert}}{2\sqrt{\vert \mbox{\sbi p}
\vert^{2}+m^{2}}}\eta^{3}
  \approx (2\pi)^{-3} \int_{^*\R^{3}}
  \frac{e^{i\mbox{\sbi p}\mbox{\sbi x}}e^{-\sqrt{\vert \mbox{\sbi p}\vert^{2}+
   m^{2}}\vert x_{0}\vert}}{2\sqrt{\vert \mbox{\sbi p}\vert^{2}+
   m^{2}}} d\mbox{\bi p}.$$
\end{lem}
\begin{proof}
 For
$$ f(x,\mbox{\bi p})=\frac{e^{i\mbox{\sbi p}\mbox{\sbi x}}
e^{-\sqrt{\vert\mbox{\sbi p}\vert^{2}+m^{2}}\vert x_{0}\vert
}}{\sqrt{\vert \mbox{\sbi p}\vert^{2}+m^{2}}}$$ calculate
\begin{multline*}
\frac{\partial}{\partial p_{\mu}}f(x,\mbox{\bi p})=\\ -
\frac{e^{i\mbox{\sbi p}\mbox{\sbi x}}e^{-\sqrt{\vert\mbox{\sbi
p}\vert ^{2} + m^{2}}\vert x_{0}\vert }p_{\mu}}{\sqrt{(\vert
\mbox{\sbi p}\vert^{2}+m^{2})^{3}}}
 +\left(ix_{\mu}-\frac{\vert x_{0}\vert p_{\mu}}{\sqrt{\vert
 \mbox{\sbi p}\vert^{2}+m^{2}}}\right)
 \frac{e^{i\mbox{\sbi p}\mbox{\sbi x}}e^{-\sqrt{\vert\mbox{\sbi p}
 \vert^{2}+m^{2}}\vert x_{0}\vert}}{\sqrt{\vert\mbox{\sbi p}
 \vert^{2}+m^{2}}}.\end{multline*}
and estimate
$$\left|\frac{\partial}{\partial p_{\mu}}f(x,\mbox{\bi p})
\right|\leq e^{-\sqrt{\vert\mbox{\sbi p}\vert^{2}+m^{2}}\vert
x_{0}\vert} \left(\frac{\vert x_{0}\vert +|x_{\mu}|}{\sqrt{\vert
 \mbox{\sbi p}\vert^{2}+m^{2}}}+\frac{1}{\vert \mbox{\sbi p}
 \vert^{2}+m^{2}}\right) .$$
 Therefore the variation of $f(x,\mbox{\bi p})$ on
$\displaystyle \prod_{\mu =1}^{3}[p_{\mu}-\eta /2, p_{\mu}+\eta/2]$
is smaller than
$$3\frac{\sqrt{\pi}}{N} e^{-\sqrt{\vert \mbox{\sbi p}\vert^{2}+
 m^{2}}\vert x_{0}\vert}\left(\frac{\vert x_{0}\vert +|x_{\mu}|}{
 \sqrt{\vert \mbox{\sbi p}\vert^{2}+m^{2}}}+\frac{1}{\vert
 \mbox{\sbi p}\vert^{2}+m^{2}}\right) .$$
This shows that
$$\left|\sum_{\mbox{\sbi p}\in\tilde{\Gamma}^{3}} f(x,\mbox{\bi p})
\eta^{3}-\int_{[-\sqrt{\pi}M, \sqrt{\pi}M]^{3}} f(x,\mbox{\bi p})
d\mbox{\bi p} \right|$$
$$\leq 3\frac{\sqrt{\pi}}{N}\int_{[-\sqrt{\pi}M,\sqrt{\pi}M]^{3}}
 \frac{(\vert x_{0}\vert +|x_{\mu}|) e^{-\sqrt{\vert \mbox{\sbi p}
 \vert^{2}+m^{2}}\vert x_{0}\vert}}{\sqrt{\vert \mbox{\sbi p}
 \vert^{2}+m^{2}}} d\mbox{\bi p}$$
$$ +3\frac{\sqrt{\pi}}{N} \int_{[-\sqrt{\pi}M, \sqrt{\pi}M]^{3}}
\frac{e^{-\sqrt{\vert \mbox{\sbi p}\vert^{2}+m^{2}}\vert x_{0}
\vert}}{\vert\mbox{\sbi p}\vert^{2}+m^{2}} d\mbox{\bi p}$$
$$\leq 3\frac{\sqrt{\pi}}{N}(\vert x_{0}|+|x_{\mu}|)\vert e^{-m\vert x_{0}
\vert /\sqrt{2}}\int_{[-\sqrt{\pi}M, \sqrt{\pi}M]^{3}}
\frac{e^{-\vert \mbox{\sbi p}\vert \vert x_{0}\vert
/\sqrt{2}}}{\sqrt{\vert\mbox{\sbi p}\vert^{2}+m^{2}}}d\mbox{\bi p}$$
$$+3\frac{\sqrt{\pi}}{N} \int_{[-\sqrt{\pi}M, \sqrt{\pi}M]^{3}}
\frac{e^{-\sqrt{\vert \mbox{\sbi p}\vert^{2}+ m^{2}}\vert x_{0}
\vert}}{\vert\mbox{\sbi p}\vert^{2}+m^{2}}d\mbox{\bi p} \approx 0.$$
Since
$$\int_{\vert \mbox{\sbi p}\vert\geq\sqrt{\pi}M}f(x,\mbox{\bi p})
 d\mbox{\bi p} \approx 0,$$
this lemma is proved.
\end{proof}
By combining Lemmas 3.4 and 3.5 we arrive at the main result of this section.
\begin{thm}
  Assume that $M,N \in \,^*\N$ are infinitely large numbers. If
$x,y \in \Gamma^4$ is finite and $|x_0-y_0|$ is not infinitesimal, then the
lattice sum (\ref{K-inverse}) is infinitesimally close to
the expected integral \begin{align}
\label{value-lattice-sum}
K^{-1}(x,y)&=(2\pi)^{-4}\sum_{p\in\tilde{\Gamma}^{4}}e^{ip(x-y)}
\left(\sum_{\mu=0}^{3} \frac{2-2\cos p_{\mu}\Delta}{\Delta
^{2}}+m^{2}\right)^{-1}\eta^{4} \nonumber\\
&\approx (2\pi)^{-3} \int_{^*\R^{3}} \frac{e^{i\mbox{\sbi p}(\mbox{\sbi x} - \mbox{\sbi y})} e^{-\sqrt{\vert \mbox{\sbi p}\vert^{2} + m^{2}} \vert x_{0} - y_{0}\vert }}{2\sqrt{\vert \mbox{\sbi p}\vert^{2} + m^{2}}} d\mbox{\bi p}
\end{align}
\end{thm}
\medskip

According to this result, the continuum limit of $K^{-1}(x, y)$
is the two point Schwinger function $S_{m}(x - y)$ of the free
neutral scalar field of mass $m$.
In fact, if $x$ and $y$ are standard real number and $x_{0} \neq  y_{0}$ then
$${}^{*}S_{m}(x - y) = \int_{{}^{*}\R^{3}} \frac{e^{i\mbox{\sbi p}(\mbox{\sbi x} - \mbox{\sbi y})} e^{-\sqrt{\vert \mbox{\sbi p}\vert^{2} + m^{2}} \vert x_{0} - y_{0}\vert }}{2\sqrt{\vert \mbox{\sbi p}\vert ^{2} + m^{2}}} d\mbox{\bi p}$$
$$      = \int _{ \R^{3}} \frac{e^{i\mbox{\sbi p}(\mbox{\sbi x} - \mbox{\sbi y})} e^{-\sqrt{\vert \mbox{\sbi p}\vert ^{2} + m^{2}} \vert x_{0} - y_{0}\vert }}{2\sqrt{\vert \mbox{\sbi p}\vert ^{2} + m^{2}}} d\mbox{\bi p} = S_{m}(x - y).$$
This follows from the transfer principle of nonstandard analysis and
means that integrations in both standard universe and nonstandard
universe coincide.  For finite $x, y \in  \Gamma $ such that
$x_{0} - y_{0}$ is not infinitesimal, we have
$${\mathcal S}_{m}(x - y) = K^{-1}(x, y) \approx {}^{*}S_{m}(x - y) \approx {}^{*}S_{m}({\rm st \, }x - {\rm st \, }y) = S_{m}({\rm st \, }x - {\rm st \, }y),$$
where ${\rm st \, }x$ is the standard part of $x$, i.e., the unique standard
real number infinitesimally close to $x$.
The two point Wightman  function thus is
\begin{align} \label{2pWf}
 &\lim_{\epsilon \rightarrow +0}
S_{m}(i(x_{0}-y_{0})+\epsilon, \mbox{\bi x}- \mbox{\bi y})
\nonumber \\
 &= \lim_{\epsilon\rightarrow+0}D_{m}^{(-)}(x_{0}-y_{0}-i\epsilon,
 \mbox{\bi x}-\mbox{\bi y})=D_{m}^{(-)}(x_{0}-y_{0},
 \mbox{\bi x}-\mbox{\bi y})\nonumber\\
&=\lim_{\epsilon\rightarrow+0}(2\pi)^{-3} \int_{^*\R^{3}}
\frac{e^{i\mbox{\sbi p}(\mbox{\sbi x}-\mbox{\sbi
y})}e^{-i\sqrt{\vert \mbox{\sbi p}
\vert^{2}+m^{2}}(x_{0}-y_{0}-i\epsilon)}}{2\sqrt{\vert\mbox{\sbi
p}\vert^{2}+m^{2}}} d\mbox{\bi p}.\end{align}

\section{Convergence of the lattice approximation of the two point Schwinger function for the free Dirac field}
We denote $\Psi(x)=(\Psi_{1}(x), \ldots, \Psi_{4}(x))^{T},$ and
recall the notation introduced in Section 2.
The discrete version of the Dirac operator then is \beq \label{FDop}
\sum_{\mu=0}^{3}\gamma^{E}_{\mu}\nabla_{\mu}+\tilde{m} .
 \eeq
The kernel of its inverse will be the lattice form of the
$2$-point Schwinger function for the free Dirac field
(compare (\ref{2pfermfunc})). Naturally, we determine this
inverse in analogy to the continuum case and use lattice
Fourier transformation instead of the standard Fourier
transformation.

 The lattice Fourier transformation transforms $\gamma ^{E}_{j}\nabla
_{\mu }\Psi (x)$  into
$$\left(\begin{array}{ccccc} 0&-i\sigma_{j}\\{} i\sigma_{j}&0\end{array} \right)
 \left( \begin{array}{ccc}(e^{ip_{j}\Delta}-1)/\Delta &0\\{} 0&(1-e^{-ip_{j}\Delta})/
 \Delta \end{array} \right) \tilde{\Psi}(p)$$
$$=\left(\begin{array}{ccc} 0&\sigma_{j}q_{j}\\{}-\sigma_{j}\bar{q}_{j}&0\end{array}
 \right) \tilde{\Psi}(p)$$
if $j = 1, 2, 3,$ with $q_{j}=- i(1-e^{-ip_{j}\Delta})/\Delta $,
respectively into
$$\left(\begin{array}{ccccc}\sigma_{0}&0\\{}0&-\sigma_{0}\end{array} \right)\left(
 \begin{array}{ccc} (e^{ip_{0}\Delta}-1)/\Delta &0\\{} 0&(1-e^{-ip_{0}\Delta })/
 \Delta \end{array} \right) \tilde{\Psi}(p)$$
$$=\left(\begin{array}{ccc}-i\sigma_{0}q_{0}&0\\{} 0&i\sigma_{0}\bar{q}_{0}
\end{array} \right) \tilde{\Psi}(p).$$
if $j=0$. Thus the Dirac operator
$$\left[\sum_{\mu=0}^{3}\gamma^{E}_{\mu}\nabla_{\mu}+\tilde{m}
\right] \Psi (x)$$
is transformed into
$$ \left(\begin{array}{ccc} i\bar{q}_{0}+\tilde{m} &\mbox{\bi \char'33 }\cdot
\mbox{\bi q}\\{} -\mbox{\bi \char'33 }\cdot \overline{\mbox{\bi
q}}&-iq_{0}+ \tilde{m} \end{array} \right)  \tilde{\Psi}(p).$$ In
order to calculate the inverse of the Dirac operator, in analogy to
the continuum case, we calculate first the inverse of the second
order differential operator of which the Dirac operator is a factor.
Accordingly we determine first this second order operator.

Under lattice Fourier transformation the operator
$$\left[-\gamma^{E}_{0}\nabla^{\prime}_{0}-\sum_{j=1}^{3}\gamma^{E}_{j}\nabla_{j}
+ \tilde{m} \right] \Psi(x),$$ where
$\nabla^{\prime}_{\mu}=P_{+}\nabla^{-}_{\mu}+P_{-}\nabla^{+}_{\mu}$,
is transformed into
$$ \left(\begin{array}{ccc}-iq_{0}+\tilde{m} &-\mbox{\bi \char'33 }\cdot
\mbox{\bi q}\\{} \mbox{\bi \char'33 }\cdot \overline{\mbox{\bi q}}&
i\bar{q} _{0}+\tilde{m} \end{array} \right) \tilde{\Psi}(p).$$ In
order to calculate the composition \beq \left(\begin{array}{ccccc}
\label{sndordpde}
 i\bar{q}_{0}+\tilde{m} &
 \mbox{\bi \char'33 } \cdot \mbox{\bi q}\\
{} -\mbox{\bi \char'33 }\cdot \overline{\mbox{\bi q}}&
-iq_{0}+
\tilde{m} \end{array} \right) \left( \begin{array}{ccc} -iq_{0}+\tilde{m} &
-\mbox{\bi \char'33 }\cdot \mbox{\bi q}\\{} \mbox{\bi \char'33 }\cdot
 \overline{\mbox{\bi q}}&i\bar{q} _{0}+\tilde{m} \end{array} \right)
  \eeq
of the above two operators we introduce the abbreviations
$a=i\bar{q}_{0}+\tilde{m}$ and $ b= \mbox{\bi \char'33 }
\cdot \mbox{\bi q}$ and find
$$\left(\begin{array}{cc} a & b \\ -\bar{b} & \bar{a} \\
     \end{array}  \right)
   \left( \begin{array}{cc} \bar{a} & -b\\   \bar{b} & a \\
     \end{array}  \right) =
     \left( \begin{array}{cc}
               a \bar{a} +  b \bar{b} & 0 \\
                 0 & \bar{a}a + \bar{b}b \\
               \end{array} \right)= \begin{pmatrix}
               K & 0\\ 0& K^*
               \end{pmatrix}
$$
with
\begin{multline*}
  K=a \bar{a} +  b \bar{b}=
(\vert q_{0}\vert^{2}+\vert \mbox{\bi q}\vert^{2}+\tilde{m}^{2}
  + 2i\tilde{m}(\bar{q}_{0}-q_{0}))\sigma_{0}\\+(q_{1}\bar{q}_{2} - q_{2}
\bar{q}_{1})\sigma_{3}+(q_{2}\bar{q}_{3}-q_{3}\bar{q}_{2})\sigma_{1}+(q_{3}
\bar{q}_{1}-q_{1}\bar{q}_{3})\sigma_{2}.
\end{multline*}
We decompose the matrix $K$ into its Hermitian and anti-Hermitian
part
$$K=D + 2i E,\quad K^*=D-2iE$$ where
\begin{align*}
D&=(\vert q_{0}\vert^{2}+\vert \mbox{\bi q}\vert^{2}+ \tilde{m}^{2}
+ 2i\tilde{m} (\bar{q}_{0}-q_{0}))\sigma_{0},\\
E&=({\rm Im\,}q_{2}\bar{q}_{3})\sigma_{1}+({\rm Im\,}q_{3}
\bar{q}_{1})\sigma_{2}+({\rm Im\,}q_{1}\bar{q}_{2})\sigma_{3}.
\end{align*}
 Observe that
\begin{align*}
  i(\bar{q} _{0}-q_{0}) &=  2{\rm Im \, }q_{0} = 2(1 - \cos p_{0}\Delta )/\Delta
 \geq  0, \\
  q_{j}\bar{q} _{k} - q_{k}\bar{q} _{j} &= 2i{\rm Im \, }(q_{j}\bar{q} _{k}).
  \end{align*}
Next we calculate the eigenvalues of the Hermitian matrix $E$ from
the equation
$$ \det (E - \lambda \sigma_{0})=\lambda^{2} - ({\rm Im \, }q_{2}\bar{q}_{3})^{2}-
({\rm Im \, }q_{3}\bar{q}_{1})^{2} - ({\rm Im \, }q_{1}\bar{q}_{2})^{2} = 0$$
and find
$$\lambda=\pm \sqrt{({\rm Im \,}q_{2}\bar{q}_{3})^{2}+({\rm Im \,}q_{3}
\bar{q}_{1})^{2}+({\rm Im \,}q_{1}\bar{q}_{2})^{2}}=\pm \rho .$$
There exist orthonormal eigenvectors $x_{\pm }$ of $E$, i.e., $E
x_{\pm}=\pm \rho x_{\pm }$, which are also eigenvectors of $D$ and
thus of $K$:
$$D x_{\pm}=(\vert q_{0}\vert^{2}+\vert \mbox{\bi q}\vert^{2}+ \tilde{m}^{2}
+ 2i\tilde{m} (\bar{q}_{0}-q_{0}))x_{\pm}=\kappa x_{\pm}$$
and
$$Kx_{\pm}=(D+2i E)x_{\pm}=(\kappa \pm 2i\rho)x_{\pm}.$$

It follows, for any $\alpha_{\pm} \in \C$, that $K(\alpha_+
x_+ +\alpha_- x_-)=\alpha_+(\kappa +2i\rho)x_+
+\alpha_-(\kappa -2i\rho)x_-$ and therefore $\norm{K x}^2
=(\kappa^2 + 4\rho^2)\norm{x}^2$ for all vectors $x$.
Therefore $(\kappa^2 +4\rho^2)^{-\frac{1}{2} }K$ is a
unitary matrix. It also follows that $(\kappa^2
+4\rho^2)^{-\frac{1}{2}} K^*$ is a unitary matrix too and
thus the following relations hold:
$$ \sqrt{\kappa^{2}+4\rho^{2}} K^{-1}=K ^{*} /\sqrt{\kappa^{2}+4\rho ^{2}}, \
 K^{-1}=K^{*} /(\kappa^{2}+4\rho^{2}).$$
$$ \sqrt{\kappa^{2}+4\rho^{2}}K^{*-1}=K /\sqrt{\kappa^{2}+4\rho^{2}},
\ K^{*-1}=K /(\kappa^{2}+4\rho^{2}).$$

Now it is straightforward to calculate the inverse of the product
operator (\ref{sndordpde}):  $$
\left(\begin{array}{ccccc}-iq_{0}+\tilde{m}&-\mbox{\bi\char'33}
\cdot \mbox{\bi q}\\{}\mbox{\bi \char'33}\cdot\overline{\mbox{\bi
q}}&i\bar{q}_{0}+\tilde{m} \end{array}\right)^{-1}\left(
\begin{array}{ccc} i\bar{q}_{0}+\tilde{m} &\mbox{\bi\char'33
}\cdot \mbox{\bi q}\\{} -\mbox{\bi \char'33}\cdot
\overline{\mbox{\bi q}}&-iq_{0}+\tilde{m}\end{array}\right)^{-1}$$
$$=\left(\begin{array}{ccccc} K^{-1}&0\\{}0&K^{*-1}\end{array}
\right)  = \frac{1}{\kappa
^{2}+4\rho^{2}}\left(\begin{array}{ccc} K^{*}&0\\{}
0&K\end{array} \right),$$ from which the inverse of the
Dirac operator in lattice Fourier transformed form is
easily calculated as
\begin{multline} \label{Diracop-inverse}
\left(\begin{array}{ccccccc}
i\bar{q}_{0}+\tilde{m}&\mbox{\bi \char'33}\cdot \mbox{\bi
q}\\{} -\mbox{\bi \char'33}\cdot \overline{\mbox{\bi
q}}&-iq_{0}+\tilde{m} \end{array}\right)^{-1}=\\
\frac{1}{\kappa^{2}+4\rho^{2}} \left(\begin{array}{ccccc}
-iq_{0}+\tilde{m} &-\mbox{\bi \char'33}\cdot \mbox{\bi
q}\\{} \mbox{\bi \char'33 }\cdot \overline{\mbox{\bi
q}}&i\bar{q} _{0}+\tilde{m} \end{array}
\right)\left(\begin{array}{ccc} K^{*}&0\\{} 0&K\end{array}
\right).\end{multline} In the rest of this section we are going to
calculate the lattice Fourier transform of this identity
and will show that the continuum limit of \beq
\label{DiracLsum} (2\pi)^{-4}\sum_{p\in
\tilde{\Gamma}^{4}}\frac{e^{ixp}}{\kappa^{2}+4\rho^{2}}
 \left(\begin{array}{ccccc} \tilde{m} &-\mbox{\bi \char'33}\cdot
 \mbox{\bi q}\\{}\mbox{\bi \char'33}\cdot \overline{\mbox{\bi q}}&
 \tilde{m} \end{array} \right) \left(\begin{array}{ccc} K^{*}&0\\
 {}0&K\end{array} \right)\eta^{4}\eeq
equals the well known integral representation of the
two-point function. As in the scalar case, the four
dimensional lattice sum is evaluated successively,
beginning with the sum $\sum_{p_0 \in \tilde{\Gamma}}$.
After some preparations, a succession of lemmas will
prepare the final result of this section, Theorem
\ref{prop4.10}.

In order to calculate the matrix
$$(\kappa^2 +4\rho^2)^{-1}K=(\kappa^2 +4\rho^2)^{-1}
(\kappa \sigma_0 + 2iE)$$
 we determine first the factor
\[\frac{\kappa}{\kappa^{2}+4\rho^{2}}=\frac{1}{2}\left(\frac{1}
{\kappa +2i\rho}+\frac{1}{\kappa -2i\rho}\right) \] and expand
$\kappa=\vert q_{0}\vert^{2}+\vert\mbox{\bi q}\vert^{2}+
\tilde{m}^{2} +
2i\tilde{m}(\bar{q}_{0}-q_{0})=(1-2\tilde{m}\Delta)(2-2\cos
p_{0}\Delta)/\Delta^{2}+\vert\mbox{\bi q}\vert^{2}+\tilde{m}^{2}$.

Next we prepare the evaluation of the sum
$$\frac{1}{1-2\Tilde{m}\Delta}\sum_{p \in \tilde{\Gamma}} \frac{e^{ixp}}
{(2 - 2\cos p\Delta)/\Delta^{2}+ B_{\pm}^{2}} \eta ,$$ $$
 B_{\pm}^{2}=\frac{A^{2}\pm i\rho}{1-2\Tilde{m}\Delta},\quad
 A^{2}=\vert
\mbox{\bi q}\vert^{2}+\tilde{m}^{2}$$ with the help of
Proposition \ref{prop:1dlsum} by the following lemma.
\begin{lem} \label{lem4.1}
For the quantities introduced above these statements hold:
 a) $0\leq\rho\leq\vert \mbox{\bi q}\vert^{2}/\sqrt{2}$, $0\leq\pm
  \arg B_{\pm}\leq\pi/6$;\\ b)  if $\vert \mbox{\bi p}\vert  \leq  M_{0} = \sqrt{M}$,
then $\rho\leq\sqrt{\pi}\vert \mbox{\bi q}\vert
^{2}/(\sqrt{2}M_{0})$ and $\arg B_{\pm}\approx 0$ if $M$ is an infinitely large number.
\end{lem}
\begin{proof} Note that
\begin{multline*}{}
q_{k}\bar{q}_{j}\Delta^{2}=(e^{ip_{j}\Delta}-1)(e^{-ip_{k}\Delta}-1)
=e^{i(p_{j}-p_{k})\Delta}-e^{ip_{j}\Delta}-e^{-ip_{k}\Delta}+1,\\
{\rm Im\,}q_{k}\bar{q}_{j}\Delta^{2}=\sin(p_{j}-p_{k})\Delta + \sin p_{j}\Delta
+ \sin p_{k}\Delta\\
   =\sin p_{j}\Delta \cos p_{k}\Delta- \cos p_{j}\Delta\sin p_{k}\Delta
 - \sin p_{j}\Delta+ \sin p_{k}\Delta\\
 =\sin p_{k}\Delta(1-\cos p_{j}\Delta)-\sin p_{j}\Delta
(1 - \cos p_{k}\Delta),\quad {\rm and}\\
({\rm Im \,}q_{k}\bar{q} _{j}\Delta^{2})^{2} =
 \sin ^{2} p_{k}\Delta(1 - \cos p_{j}\Delta)^{2}\\ -2\sin p_{k}\Delta
  (1 -\cos p_{j}\Delta) \sin p_{j}\Delta(1 - \cos p_{k}\Delta)
   + \sin^{2} p_{j}\Delta(1 - \cos p_{k}\Delta )^{2},
  \end{multline*}
\begin{align*}{}&
\rho^{2}\Delta^{4}=\\
 & [({\rm Im\,}q_{2}\bar{q}_{3})^{2} + ({\rm Im
\, }q_{3} \bar{q} _{1})^{2} + ({\rm Im \, }q_{1}\bar{q}
_{2})^{2}]\Delta^{4} =
\sin ^{2} p_{2}\Delta  (1 - \cos p_{3}\Delta )^{2}\\
   &- 2\sin p_{2}\Delta
 (1 - \cos p_{3}\Delta ) \sin p_{3}\Delta  (1 - \cos p_{2}\Delta ) +
 \sin ^{2} p_{3}\Delta  (1 - \cos p_{2}\Delta )^{2}\\
 & + \sin ^{2} p_{3}\Delta  (1 - \cos p_{1}\Delta )^{2} -2\sin p_{3}\Delta
 (1 - \cos p_{1}\Delta ) \sin p_{1}\Delta  (1 - \cos p_{3}\Delta )\\
&+\sin^{2} p_{1}\Delta(1- \cos
p_{3}\Delta)^{2}+\sin^{2}p_{1}\Delta(1-\cos p_{2}
\Delta)^{2}\\
&     -2\sin p_{1}\Delta(1-\cos p_{2}\Delta)\sin p_{2}\Delta(1-\cos p_{1}\Delta )
 + \sin^{2}p_{2}\Delta(1-\cos p_{1}\Delta)^{2}\\
 & =(1-\cos p_{1}\Delta)^{2}[\sin^{2} p_{3}\Delta + \sin^{2} p_{2}\Delta]+
(1-\cos p_{2}\Delta )^{2}[\sin^{2} p_{3}\Delta +\\ & +\sin^{2} p_{1}\Delta]
+(1-\cos p_{3}\Delta)^{2}[\sin^{2} p_{2}\Delta+ \sin {2} p_{1}\Delta]\\
&-2\sin p_{2}\Delta \sin p_{3}\Delta(1-\cos p_{3}\Delta)(1-\cos p_{2}\Delta)\\
&-2\sin p_{3}\Delta\sin p_{1}\Delta(1-\cos p_{1}\Delta)(1-\cos p_{3}\Delta)\\
&-2\sin p_{1}\Delta\sin p_{2}\Delta(1-\cos p_{2}\Delta)(1-\cos
p_{1}\Delta).
\end{align*}
Since $\vert \sin p_{\mu }\Delta \vert  \leq 1$, we have
\begin{align*}
\rho^{2}\Delta^{4} & \leq 2 \sum_{j=1}^{3}(1-\cos
p_{j}\Delta)^{2}+2(1-\cos p_{3}\Delta)(1-\cos p_{2}
\Delta)\\
&+2(1-\cos p_{1}\Delta)(1-\cos p_{3}\Delta)+2(1-\cos p_{2}\Delta)(1-\cos p_{1}
\Delta)\\
&\leq \frac{1}{2}\left(\sum_{j=1}^{3}(2-2\cos
p_{j}\Delta)\right)^{2}=\frac{1}{2} (\vert \mbox{\bi q}\vert
^{2})^{2} \Delta^{4}.
\end{align*}
Thus we have $\rho  \leq  \vert \mbox{\bi q}\vert ^{2}/\sqrt{2}$.  Since $0 \leq  \pm  \arg B^{2}_{\pm } $,
$$0\leq\pm\arg B_{\pm}^{2}\leq\tan^{-1}\sqrt{2}\leq\pi/3,\ 0\leq
\pm\arg B_{\pm}\leq\pi/6.$$ If $\vert \mbox{\bi p}\vert  \leq
M_{0}$, then $\vert \sin p_{\mu }\Delta \vert  \leq \vert p_{\mu
}\Delta \vert  \leq  \sqrt{\pi }M_{0}^{-1}$, and therefore
$\rho^{2}\Delta^{4} \leq \pi (\vert \mbox{\bi q}\vert ^{2})^{2}
\Delta^{4}/(2M_0^2)$
 and
$\rho \leq \sqrt{\pi }\vert \mbox{\bi q}\vert
^{2}/(\sqrt{2}M_{0})$. Hence $\tan({\arg B_{\pm}^2}) \leq \frac{1}{M_0}\sqrt{\pi/2}$
and thus $\arg B_{\pm} \approx 0$ if $M$ is infinitely large.
\end{proof}
By Lemma \ref{lem4.1}, $B_{\pm }$ satisfies the conditions
for the constant $B$ in Proposition \ref{prop:1dlsum}, so
this proposition applies for the present case and yields
\begin{prop} \label{prop4.2}
For the quantities $B_{\pm}$ introduced above we have
  $B_{\pm} \neq 0, \vert \arg B_{\pm}\vert \leq \pi/6$ and, for all
 $x \in \Gamma$,
$$ \sum_{p \in \tilde{\Gamma}}\frac{e^{ixp}}{(2-2\cos p\Delta)/\Delta^{2}+
B_{\pm}^{2}} \eta=$$
$$ = \frac{2\pi(1+\Delta B_{\pm}[\sqrt{4+\Delta^{2}B_{\pm}^{2}}/2 +
\Delta B_{\pm}/2])^{-\vert
 x\vert /\Delta}}{B_{\pm}\sqrt{4+\Delta^{2}B_{\pm}^{2}}}.$$
If $M, N \in {}^{*}\N$ are infinitely large numbers and $\delta
 = \Delta B_{\pm}[\sqrt{4 +\Delta^{2}B_{\pm}^{2}}/2 + \Delta
 B_{\pm}/2]$ is infinitesimal, then, for some
$e_{*} \approx e$, the above sum equals
$$ \frac{2\pi e_{*}^{-B_{\pm}[\sqrt{4+\Delta^{2}B_{\pm}^{2}}/2 +\Delta
B_{\pm}]|x|}} {B_{\pm}\sqrt{4+\Delta^{2}B_{\pm}^{2}}}$$
which is near $2\pi e^{-B_{\pm}|x|}/2B_{\pm}$ and
less than $2\pi \vert 2^{-B|x|}/2B_{\pm}\vert $.\end{prop}

Our sum is evaluated further and estimated in the next proposition.
\begin{prop}\label{prop4.3}
 Let $M, N \in {}^{*}\N$ be infinitely large numbers and
$M_{0}=\sqrt{M}$.  If $\vert \mbox{\bi p}\vert \leq  M_{0}$, then for $x_0 \in \Gamma$
\begin{align} \label{eq4.1}
 &(2\pi)^{-4} \sum_{p_{0}\in \tilde{\Gamma}}
 \frac{e^{ixp}}{(2 - 2\cos p_{0} \Delta)/\Delta^{2} + B_{\pm}^{2}} \eta  \nonumber \\
      = &(2\pi)^{-3} \frac{e^{i\mbox{\sbi p}\mbox{\sbi x}}
e_{*}(\mbox{\bi p})^{-B_{\pm}[\sqrt{4 + \Delta^{2}B_{\pm}^{2}}/2 +
\Delta B_{\pm }/2]|x_{0}|}}{B_{\pm}\sqrt{4 + \Delta
^{2}B_{\pm}^{2}}},
\end{align}
and $e_{*}(\mbox{\bi p}) \approx e$.
If $\vert \mbox{\bi p}\vert  \geq  M_{0}$ then
\begin{align} \label{eq4.2}
  & \left|(2\pi)^{-4} \sum _{p_{0}\in \tilde{\Gamma}}
   \frac{e^{ixp}}{(2-2\cos p\Delta)/\Delta^{2}+B_{\pm}^{2}} \eta \right|\nonumber \\
     \leq & (2\pi )^{-3} 2^{-2M_{0}\vert x_{0}\vert /\pi }\frac{1}{4M_{0}/\pi }.
 \end{align}
\end{prop}
\begin{proof}
If $\vert \mbox{\bi p}\vert  \leq  M_{0}$, then $\rho  \leq  \sqrt{\pi }\vert
\mbox{\bi q}\vert ^{2}/(\sqrt{2}M_{0})$ and
\begin{multline*}
\vert B_{\pm}^{2}\vert \leq \\ \frac{A^{2} + \sqrt{2\pi}\vert
\mbox{\bi q}\vert^{2}/M_{0}}{1+ \Delta} \leq  A^{2} \frac{1 +
\sqrt{2\pi }/M_{0}} {1 + \Delta} \leq \sqrt{M_{0}^{2}+\tilde{m}^{2}}
\frac{1+\sqrt{2\pi}/M_{0}} {1 + \Delta}.\end{multline*}

This shows that $\Delta
B_{\pm }$ and $\delta=\Delta B_{\pm }[\sqrt{4+\Delta^{2}B_{\pm
}^{2}}/2+\Delta B_{\pm}/2]$ are infinitesimal.  Application of Proposition
\ref{prop4.2} implies (\ref{eq4.1}).

Apply Lemma \ref{lem4.1} to $\arg B_{\pm}=\theta_{\pm}$.
This gives $\vert \theta_{\pm}\vert \leq \pi/6$. For
$\phi_{\pm}=\arg \sqrt{4 + \Delta ^{2}B_{\pm }^{2}}$ we get
$0 \leq\pm\phi_{\pm} < \pm \theta_{\pm}$ and, since
$$\vert B_{\pm}^{2}\vert\geq\frac{A^{2}}{1-2\Tilde{m}\Delta}, \  \vert 4 +
\Delta^{2}B_{\pm}^{2}\vert\geq
4+\Delta^{2}\frac{A^{2}}{1-2\Tilde{m}\Delta },$$ it follows
$$\vert B_{\pm}\vert\vert\sqrt{4+\Delta^{2}B_{\pm}^{2}}\vert\geq
\frac{A\sqrt{4+\Delta^{2}A^{2}}}{1-2\Tilde{m}\Delta},$$
$${\rm Re\,}B_{\pm}\sqrt{4+\Delta^{2}B_{\pm}^{2}}=\vert B_{\pm}
\vert \vert \sqrt{4+\Delta^{2}B_{\pm}^{2}}\vert{\rm Re\,}e^{i\theta
_{\pm}}e^{i\phi_{\pm}}$$
$$\geq\frac{1}{2}\vert B_{\pm}\vert \vert \sqrt{4+\Delta^{2}B_{\pm}
^{2}}\vert\geq\frac{1}{2(1-2\Tilde{m}\Delta)} A
\sqrt{4+\Delta^{2}A^{2}}.$$ Since (recall that $z_{\pm}$ is
defined by Eq. (\ref{zplusminus}) with $B$ replaced by $B_{\pm}$)
$$\vert z_{+}\vert\geq{\rm Re \,}z_{+}=\frac{2+{\rm Re \,}\Delta^{2}
B_{\pm}^{2}+{\rm Re \,}\Delta B_{\pm}\sqrt{4+\Delta^{2}B_{\pm
}^{2}}}{2}$$
$$\geq 1+(1/2)[\Delta^{2}A^{2}+(1/2)\Delta A\sqrt{4+\Delta^{2}A^{2}}]
/(1-2\Tilde{m}\Delta),$$ we have
\begin{multline} \label{eq3.3}
  \frac{\vert z_{+}^{-\vert x_{0}\vert /\Delta }\vert }{\vert B_{\pm }\sqrt{4 +
  \Delta^{2}B_{\pm}^{2}}\vert}\leq \\ \frac{\{1+(1/4)\Delta A[\sqrt{4+
  \Delta^{2}A^{2}}+2\Delta A]/(1-2\Tilde{m}\Delta)\}^{-\vert x_{0}\vert
  /\Delta }}{A\sqrt{4+\Delta^{2}A^{2}}/(1-2\Tilde{m}\Delta)}.\end{multline}
If $\vert \mbox{\bi p}\vert=M_{0}$, then $\delta=(1/4)\Delta
A[\sqrt{4+\Delta^{2}A^{2}}+2\Delta A]/(1-2\Tilde{m}\Delta )$ is
infinitesimal, and the right hand side of Eq. (\ref{eq3.3}) is:
$$\frac{\{(1+\delta)^{1/\delta}\}^{-\delta \vert x_{0}\vert
/\Delta}}{A\sqrt{4+\Delta^{2}A^{2}}}=\frac{e_{*}^{-\{ (1/4)A
[\sqrt{4+\Delta^{2}A^{2}}+2\Delta A]/(1-2\Tilde{m}\Delta)\}\vert
x_{0}\vert }}{A\sqrt{4+\Delta^{2}A^{2}}}$$
$$=\frac{(e_{**}^{1/(1-2\Tilde{m}\Delta)})^{-\{(1/2)A\} \vert x_{0}\vert}}
{2A} \leq  \frac{2^{-\{(1/2)A\} \vert x_{0}\vert }}{2A}
\leq 2^{-2M_{0}\vert x_{0}\vert
/\pi}\frac{1}{4M_{0}/\pi},$$ where we used again the fact
that
$$\vert\mbox{\bi q}\vert^{2}=\sum_{\mu=1}^{3} \frac{2-2\cos
p_{\mu}\Delta}{\Delta^{2}}\geq  4/\pi^{2} \sum_{\mu=1}^{3} \vert
p_{\mu}\vert^{2}=4/\pi^{2} \vert \mbox{\bi p}\vert^{2} \geq
4M_{0}^{2}/\pi^{2}.$$ Since the right hand side of Eq. (\ref{eq3.3})
is a decreasing function of $A$, we estimate, for $\vert \mbox{\bi
p}\vert \geq M_{0}$, as follows:
$$\frac{\vert z_{+}^{-\vert x_{0}\vert /\Delta}\vert}{\vert
B_{\pm}\sqrt{4+\Delta^{2}B_{\pm}^{2}}\vert} \leq 2^{-2M_{0}\vert
x_{0}\vert /\pi }\frac{1}{4M_{0}/\pi }.$$ This proves (\ref{eq4.2}).
\end{proof}
After these preparations, in a sequence of lemmas, the
evaluation of the four dimensional lattice sum for the
Dirac field is done successively. The first step is:

\begin{lem}\label{lem4.3}
For infinitely large numbers $M, N \in{}^{*}\N$, if $x_{0}\in \Gamma $ is not infinitesimal, the four dimensional
lattice sum is approximated by a three dimensional one:
$$ (2\pi)^{-4}\sum_{p\in \tilde{\Gamma}^{4}}\frac{e^{ixp}}{(2-2\cos p_{0}
\Delta)/\Delta^{2} + B_{\pm}^{2}} \eta ^{4} $$ \nopagebreak
$$ \approx (2\pi)^{-3} \sum_{\mbox{\sbi p}\in \tilde{\Gamma}^{3}}
\frac{e^{i\mbox{\sbi p} \mbox{\sbi x}}e^{-\sqrt{\vert \mbox{\sbi
p}\vert^{2}+\tilde{m}^{2}}\vert x_{0}\vert}}{2\sqrt{\vert \mbox{\sbi
p}\vert^{2}+\tilde{m}^{2}}}\eta^{3}.$$
\end{lem}
\begin{proof}
With the abbreviation  $M_{0} = \sqrt{M}$ we estimate as follows:
$$\left|(2\pi)^{-4} \sum_{\mbox{\sbi p}\in \tilde{\Gamma}^{3},\vert
\mbox{\sbi p}\vert \geq M_{0}}\sum_{p_{0}\in \tilde{\Gamma}}
\frac{e^{ixp}}{(2 - 2\cos p_{0}\Delta )/\Delta ^{2} + B_{\pm }^{2}}
\eta^{4}\right| $$
$$\leq(2\pi)^{-3} \sum_{\mbox{\sbi p}\in \tilde{\Gamma}^{3},\vert
\mbox{\sbi p}\vert\geq M_{0}}\frac{2^{-2M_{0}\vert x_{0}\vert
/\pi}}{4M_{0}/\pi}\eta^{3} \leq(2\pi)^{-3/2}
M^{3}\frac{2^{-2M_0\vert x_{0}\vert /\pi}}{4M_0/\pi} \approx 0.$$

Since
$$\left|\frac{e^{i\mbox{\sbi p}\mbox{\sbi x}} e_{*}(\mbox{\bi p})^{-B_{\pm }
[\sqrt{4 + \Delta ^{2}B_{\pm }^{2}}/2 + \Delta B_{\pm }/2]|x_{0}|}}{B_{\pm }
\sqrt{4 + \Delta ^{2}B_{\pm }^{2}}} \right| $$
$$\leq \frac{2^{-A\vert x_{0}\vert}}{2A}\leq \frac{2^{-2\vert
\mbox{\sbi p}\vert \vert x_{0}\vert /\pi}}{4\vert \mbox{\sbi
p}\vert/\pi }$$ for $\vert \mbox{\bi p}\vert  \leq  M_{0}$, for any
standard $\epsilon  > 0$, there exists a finite $M_{1}>0$ such that
$$\left|(2\pi)^{-3}\sum_{\mbox{\sbi p}\in\tilde{\Gamma}^{3},
M_{1}\leq\vert\mbox{\sbi p}\vert\leq M_{0}}\frac{e^{i\mbox{\sbi
p}\mbox{\sbi x}}e_{*}(\mbox{\bi p})^{-B_{\pm}[\sqrt{4+\Delta
^{2}B_{\pm}^{2}}/2 + \Delta B_{\pm
}/2]|x_{0}|}}{B_{\pm}\sqrt{4+\Delta ^{2}B_{\pm}^{2}}} \eta ^{3}
\right|<\epsilon $$ if $\vert x_{0}\vert $ is not infinitesimal.
This shows that
$$ \forall\; \epsilon >0\; \exists\; M_{1}\;\left|(2\pi)^{-4}
\sum_{\mbox{\sbi p}\in\tilde{\Gamma}^{3}, M_{1} \leq \vert
\mbox{\sbi p}\vert} \sum_{p_{0}\in \tilde{\Gamma}}
\frac{e^{ixp}}{(2-2\cos p_{0}\Delta)/\Delta^{2}+B_{\pm}^{2}} \eta
^{4} \right|<\epsilon .$$ We also know
$$\forall\;\epsilon >0\; \exists\; M_{1}\; \left|(2\pi)^{-3}
\sum_{\mbox{\sbi p}\in \tilde{\Gamma}^{3}, \vert \mbox{\sbi p}\vert
\geq  M_{1}}\frac{e^{-\sqrt{\vert\mbox{\sbi p}\vert^{2}+\tilde{m}
^{2}}\vert x_{0}\vert}}{\sqrt{\vert\mbox{\sbi p}\vert^{2} +
\tilde{m}^{2}}} \eta^{3} \right| < \epsilon.$$ If $\vert \mbox{\bi
p}\vert < M_{1}$ for a finite $M_{1}>0$, then it follows that
$$\rho\approx 0,\ B_{\pm} \approx A \approx \sqrt{\mbox{\bi p}^{2}+
 \tilde{m}^{2}}$$
and
$$\frac{e_{*}(\mbox{\bi p})^{-B_{\pm }[\sqrt{4+\Delta^{2}
B_{\pm}^{2}}/2+\Delta
B_{\pm}/2]|x_{0}|}}{B_{\pm}\sqrt{4+\Delta^{2} B_{\pm}^{2}}}
\approx \frac{e^{-\sqrt{\vert \mbox{\sbi p}\vert^{2}+
\tilde{m}^{2}}\vert x_{0}\vert}}{2\sqrt{\vert\mbox{\sbi
p}\vert^{2} +\tilde{m}^{2}}}.$$ Hence, for non-infinitesimal
$\vert x_{0}\vert $,  we conclude
$$(2\pi)^{-3} \sum_{\mbox{\sbi p}\in \tilde{\Gamma}^{3},\vert
\mbox{\sbi p}\vert \leq M_{1}}\frac{e^{i\mbox{\sbi p}\mbox{\sbi
x}} e_{*}(\mbox{\bi p})^{-B_{\pm
}[\sqrt{4+\Delta^{2}B_{\pm}^{2}}/2+ \Delta B_{\pm
}/2]|x_{0}|}}{B_{\pm}\sqrt{4+\Delta^{2}B_{\pm}^{2}}} \eta ^{3}$$
$$\approx (2\pi)^{-3} \sum_{\mbox{\sbi p}\in \tilde{\Gamma}^{3},
\vert \mbox{\sbi p}\vert \leq M_{1}} \frac{e^{i\mbox{\sbi
p} \mbox{\sbi x}}e^{-\sqrt{\vert \mbox{\sbi
p}\vert^{2}+\tilde{m} ^{2}}\vert
x_{0}\vert}}{2\sqrt{\vert\mbox{\sbi p}\vert^{2}+
\tilde{m}^{2}}} \eta^{3}$$
  and the proof is complete.
\end{proof}
\begin{lem} \label{lem4.4}
 Under the same condition as Lemma \ref{lem4.3} the four dimensional lattice sum
$$(2\pi)^{-4} \sum_{p\in \tilde{\Gamma}^{4}} \frac{e^{ixp}
q_{k}}{(2 - 2\cos p_{0}\Delta)/\Delta^{2} + B_{\pm}^{2}} \eta
^{4}$$
is infinitesimally close to the spatial three dimensional one
$$\approx (2\pi)^{-3} \sum_{\mbox{\sbi p}\in \tilde{\Gamma}^{3}}
\frac{e^{i\mbox{\sbi p} \mbox{\sbi x}} p_{k} e^{-\sqrt{\vert \mbox{\sbi p}
\vert ^{2} + \tilde{m} ^{2}}\vert x_{0}\vert }}{2\sqrt{\vert \mbox{\sbi p}
\vert ^{2} + \tilde{m} ^{2}}} \eta ^{3}$$
for $k = 1, 2, 3$.
\end{lem}
\begin{proof}
We can proceed in the same way as in the proof of Lemma
\ref{lem4.3}.
\end{proof}
\begin{lem}\label{lem4.5}
 Under the same condition as Lemma \ref{lem4.3}, the four
 dimensional lattice sum for the Dirac operator is reduced to a three dimensional
 one in the following way:
$$(2\pi)^{-4}\sum_{p\in \tilde{\Gamma}^{4}}\frac{e^{ixp}}{\kappa^{2}+4\rho^{2}}
 \left(\begin{array}{ccccc} \tilde{m} &-\mbox{\bi \char'33}\cdot
 \mbox{\bi q}\\{}\mbox{\bi \char'33}\cdot \overline{\mbox{\bi q}}&
 \tilde{m} \end{array} \right) \left(\begin{array}{ccc} K^{*}&0\\
 {}0&K\end{array} \right)\eta^{4} $$
$$      \approx (2\pi )^{-3} \sum _{\mbox{\sbi p}\in \tilde{\Gamma } ^{3}}
\frac{e^{i\mbox{\sbi p} \mbox{\sbi x}} e^{-\sqrt{\vert \mbox{\sbi p}\vert ^{2} +
\tilde{m} ^{2}}\vert x_{0}\vert }}{2\sqrt{\vert \mbox{\sbi p}\vert ^{2} +
\tilde{m}^{2}}} \left( \begin{array}{ccc} \tilde{m} &-\mbox{\bi \char'33 }\cdot
\mbox{\bi p}\\{} \mbox{\bi \char'33 }\cdot \overline{\mbox{\bi p}}&\tilde{m}
\end{array} \right)  \eta ^{3}.$$
\end{lem}
\begin{proof}
Note that if $\rho  \neq  0$
$$ \sum _{p_{0}\in \tilde{\Gamma}}\frac{e^{ix_0p_0}}{\kappa ^{2}+4\rho ^{2}}
 K \eta  = \sum _{p_{0}\in \tilde{\Gamma}} \frac{e^{ix_0p_0}}{\kappa^{2}+4\rho^{2}}
  (\kappa \sigma _{0} + 2iE) \eta  $$
$$ = \sum_{p_{0}\in \tilde{\Gamma} } e^{ix_0p_0} \frac{1}{2} \left(  \frac{1}
{\kappa  - 2i\rho } + \frac{1}{\kappa  + 2i\rho }\right) \sigma _{0} \eta  $$
$$+ \sum_{p_{0}\in \tilde{\Gamma}} e^{ix_0p_0} \frac{1}{2\rho } \left( \frac{1}
{\kappa  - 2i\rho } - \frac{1}{\kappa  + 2i\rho }\right) E \eta  ,$$
and
$$ \sum_{p_{0}\in \tilde{\Gamma}} \frac{e^{ix_0p_0}}{\kappa^{2}+4\rho^{2}} K \eta =
 \sum_{p_{0}\in \tilde{\Gamma}} \frac{e^{ix_0p_0}}{\kappa}\sigma_{0}\eta  $$
if $\rho  = 0$.  Since
$$ \sum_{p_{0}\in \tilde{\Gamma}}e^{ix_0p_0} \frac{1}{\kappa \pm
  2i\rho}\eta =\frac{1}{1 + \Delta}\sum_{p_{0}\in \tilde{\Gamma}}
 \frac{e^{ix_0p_0}}{(2-2\cos p_{0}\Delta)/\Delta^{2}+B_{\pm}^{2}}\eta $$
$$= \frac{2\pi}{1 + \Delta} \frac{(1+ \Delta B_{\pm}[\sqrt{4 +
\Delta^{2}B_{\pm}^{2}}/2 +\Delta B_{\pm}/2])^{|x_0|/\Delta}}{B_{\pm}
\sqrt{4 + \Delta ^{2}B_{\pm }^{2}}}$$ and $E/\rho $ is finite, this
lemma follows from Lemmas \ref{lem4.3} and \ref{lem4.4}.
\end{proof}
\begin{lem} \label{lem4.6}
Under the same condition as Lemma \ref{lem4.3} the following approximation holds:
$$(2\pi)^{-4} \sum_{p\in \tilde{\Gamma}^{4}} \frac{i\bar{q}_{0}
e^{ixp}}{(2 - 2\cos p_{0}\Delta)/\Delta^{2} + B_{\pm}^{2}} \eta^4 $$
$$\approx-\frac{x^{0}}{\vert x^{0}\vert}(2\pi)^{-3}\sum_{\mbox{
\sbi p}\in \tilde{\Gamma}^{3}} \frac{e^{i\mbox{\sbi p} \mbox{\sbi
x}}e^{-\sqrt{\vert \mbox{\sbi p}\vert^{2} + \tilde{m}^{2}}\vert
x^{0}\vert}}{2} \eta^{3}.$$
\end{lem}
\begin{proof}
$$\sum_{p_0\in\tilde{\Gamma}}\frac{i\bar{q_0}e^{ix_0p_0}}
{(2 - 2 \cos p\Delta )/\Delta^{2} + B_{\pm}^{2}}\eta =\sum_{p_0\in
\tilde{\Gamma}}\frac{\nabla^{+}e^{ix_0p_0}}{(2 - 2 \cos p\Delta )/
\Delta^{2} + B_{\pm}^{2}}\eta $$
$$=\frac{2\pi \nabla^{+} z_{-}^{x_0/\Delta}}{z_{+} - z_{-}} =
\frac{2\pi z_{-}^{x_0/\Delta}(z_{-} - 1)/\Delta}{z_{+} - z_{-}}$$
$$= \frac{2\pi z_{-}^{x_0/\Delta}}{z_{+} - z_{-}}
 \frac{\Delta ^{2}B_{\pm}^{2}-B_{\pm}\sqrt{4 + \Delta^{2}B_{\pm}^{2}}}{2}$$
if $x_0 \geq  0$, and if $x_0 < 0$, then
$$= \frac{2\pi   z_{+}^{x_0/\Delta}}{z_{+} - z_{-}} \frac{\Delta ^{2}B^{2} +
B\sqrt{4 + \Delta ^{2}B^{2}}}{2}.$$
  Now we can prove this lemma in the same way as
Lemma \ref{lem4.3}, since
$$\frac{\Delta^{2}B_{\pm}^{2} \pm \sqrt{4 + \Delta^{2}B_{\pm}^{2}}}{2\sqrt{4 + \Delta ^{2}B^{2}}} \approx \pm  \frac{1}{2}$$
for $\vert \mbox{\bi p}\vert  \leq  M_{0}$.
\end{proof}
\begin{lem} \label{lem4.7}
 Under the same condition as Lemma \ref{lem4.3} the following approximation
$$(2\pi)^{-4}\sum_{p\in \tilde{\Gamma}^{4}} \frac{e^{ixp}}{\kappa^{2}+4\rho^{2}} \left(
\begin{array}{ccccc} -iq_{0}{}&0\\{} 0&i\bar{q} _{0}\end{array} \right)
\left( \begin{array}{ccc} K^{*}&0\\{} 0&K\end{array} \right)  \eta ^{4} $$
$$ \approx \frac{x_{0}}{\vert x_{0}\vert}\gamma_{0}^{E} (2\pi)^{-3}
 \sum_{\mbox{\sbi p}\in \tilde{\Gamma}^{3}} \frac{e^{i\mbox{\sbi p}
  \mbox{\sbi x}} e^{-\sqrt{\vert \mbox{\sbi p}\vert ^{2} + \tilde{m} ^{2}}
  \vert x_{0}\vert}}{2} \eta^{3}$$
  of the four dimensional  by a three dimensional lattice sum holds.
\end{lem}
\begin{proof}
  We can prove this lemma in the same way as Lemma \ref{lem4.4}.
  \end{proof}
The combination of lemmas \ref{lem4.5} and \ref{lem4.7}
gives
\begin{lem} \label{lem4.8}
 Under the same condition as Lemma \ref{lem4.3}, the four
 dimensional lattice sum for the Dirac operator is reduced to a
 three dimensional one:
 $$(2\pi)^{-4}\sum_{p\in \tilde{\Gamma}^{4}}\frac{e^{ixp}}{\kappa^{2}+4\rho^{2}}
\left(\begin{array}{ccccc}-iq_{0}+\tilde{m} &-\mbox{\bi
\char'33 }\cdot \mbox{\bi q}\\{} \mbox{\bi \char'33 }\cdot
\overline{\mbox{\bi q}}&i\bar{q} _{0}+\tilde{m} \end{array}
\right) \left( \begin{array}{ccc} K^{*}&0\\{}
0&K\end{array} \right)  \eta ^{4} $$
$$  \approx (2\pi )^{-3} \sum _{\mbox{\sbi p}\in \tilde{\Gamma } ^{3}} \frac{e^{i\mbox{\sbi p} \mbox{\sbi x}} e^{-\sqrt{\vert \mbox{\sbi p}\vert ^{2} + \tilde{m} ^{2}}\vert x_{0}\vert }}{2\sqrt{\vert \mbox{\sbi p}\vert ^{2} + \tilde{m} ^{2}}} \left( \begin{array}{ccc} \tilde{m} &-\mbox{\bi \char'33 }\cdot \mbox{\bi p}\\{} \mbox{\bi \char'33 }\cdot \overline{\mbox{\bi p}}&\tilde{m} \end{array} \right)  \eta ^{3}$$
$$ + \frac{x_{0}}{\vert x_{0}\vert }\gamma _{0}^{E} (2\pi )^{-3}
 \sum _{\mbox{\sbi p}\in \tilde{\Gamma } ^{3}} \frac{e^{i\mbox{\sbi p}
 \mbox{\sbi x}} e^{-\sqrt{\vert \mbox{\sbi p}\vert^{2} + \tilde{m}^{2}}
 \vert x_{0}\vert }}{2} \eta^{3}.$$
\end{lem}
\begin{lem} \label{lem4.9}
Assume that $M, N \in  {}^{*}\N$ are infinitely large numbers.  If
$\mbox{\bi x}$ is finite and $\vert x_{0}\vert $ not infinitesimal, then the `spatial'
lattice sum is approximated by the expected
integral:
$$ (2\pi)^{-3} \sum_{\mbox{\sbi p}\in \tilde{\Gamma}^{3}}
\frac{e^{i\mbox{\sbi p}\mbox{\sbi x}} e^{-\sqrt{\vert \mbox{\sbi
p}\vert^{2} + \tilde{m}^{2}}\vert x_{0} \vert }}{2\sqrt{\vert
\mbox{\sbi p}\vert^{2} + \tilde{m}^{2}}} \left(\begin{array}{ccc}
\tilde{m} &-\mbox{\bi \char'33 }\cdot \mbox{\bi p}\\{}\mbox{\bi
\char'33}\cdot \overline{\mbox{\bi p}}&\tilde{m} \end{array} \right)
 \eta ^{3}$$
$$  + \frac{x_{0}}{\vert x_{0}\vert}\gamma_{0}^{E} (2\pi)^{-3}
 \sum_{\mbox{\sbi p}\in \tilde{\Gamma}^{3}} \frac{e^{i\mbox{\sbi p} \mbox{\sbi x}}
e^{-\sqrt{\vert \mbox{\sbi p}\vert^{2} + \tilde{m}^{2}}\vert x_{0}
\vert}}{2} \eta ^{3}.$$
$$ \approx (2\pi)^{-3} \int_{^*\R^{3}} \frac{e^{i\mbox{\sbi p}
\mbox{\sbi x}} e^{-\sqrt{\vert \mbox{\sbi p}\vert^{2} +
\tilde{m}^{2}}\vert x_{0} \vert}}{2\sqrt{\vert \mbox{\sbi
p}\vert^{2} + \tilde{m}^{2}}} \left(\begin{array}{ccc} \tilde{m}
&-\mbox{\bi \char'33 }\cdot \mbox{\bi p}\\{} \mbox{\bi \char'33
}\cdot \overline{\mbox{\bi p}}&\tilde{m} \end{array} \right)
  \eta ^{3} d\mbox{\sbi p}$$
$$ + \frac{x_{0}}{\vert x_{0}\vert}\gamma _{0}^{E} (2\pi)^{-3}
\int_{ ^*\R^{3}} \frac{e^{i\mbox{\sbi p} \mbox{\sbi x}}
e^{-\sqrt{\vert \mbox{\sbi p}\vert^{2} + \tilde{m}^{2}}\vert
x_{0}\vert}}{2} \eta ^{3}d \mbox{\sbi p}.$$
\end{lem}
\begin{proof}
 The proof strategy of Lemma 3.5 applies.
\end{proof}
Using the formulae ($A>0$)
$$(2\pi)^{-1}\int_{\R}\frac{e^{ipx}}{p_{0}^{2}+A^{2}} dp_{0} =
\frac{e^{i\mbox{\sbi p}\mbox{\sbi x}}e^{-A\vert x_{0}\vert}}{2A},$$
$$(2\pi)^{-1}\int_{\R}\frac{-ip_{0}e^{ipx}}{p_{0}^{2}+A^{2}}dp_{0}
 =\frac{x_0}{|x_0|} \frac{e^{i\mbox{\sbi p} \mbox{\sbi x}} e^{-A\vert x_{0}\vert }}{2}, $$
 we get the main result of this section.

\begin{thm} \label{prop4.10}
 For infinitely large numbers $M,N\in{}^{*}\N$, for all finite $x\in\Gamma^4$
 for which $\vert x_{0}\vert $ is not infinitesimal, the lattice
 Fourier transform of (\ref{Diracop-inverse}) is given by
 \begin{align}\label{eq:mainresult}
&\quad(4\pi)^{-4}\sum_{p\in \tilde{\Gamma}^4} e^{ipx} {
\begin{pmatrix}
i\bar{q_0}+\tilde{m}&\mbox{\bi \char'33 }\cdot \mbox{\bi q}\\
-\mbox{\bi \char'33 }\cdot \overline{\mbox{\bi q}}& -i q_0
+\tilde{m}
\end{pmatrix} }^{-1} \eta^4 = \nonumber \\
&=(2\pi)^{-4}\sum_{p\in \tilde{\Gamma}^{4}}
\frac{e^{ipx}}{\kappa^{2}+4\rho^{2}} \left( \begin{array}{ccccc}
-iq_{0}+\tilde{m} &-\mbox{\bi \char'33 }\cdot \mbox{\bi q}\\{}
\mbox{\bi \char'33 }\cdot \overline{\mbox{\bi q}}&i\bar{q}
_{0}+\tilde{m} \end{array} \right) \left( \begin{array}{ccc}
K^{*}&0\\{} 0&K\end{array} \right)  \eta ^{4}\nonumber \\
&\approx (2\pi)^{-4} \int_{^*\R^{4}} \frac{e^{ipx}}{p^{2}+m^{2}}
\left( \begin{array}{ccc} -ip_{0}+\tilde{m} &-\mbox{\bi \char'33
}\cdot \mbox{\bi p}\\{} \mbox{\bi \char'33 }\cdot \mbox{\bi
p}&ip_{0}+\tilde{m}
\end{array} \right) dp.\end{align}
\end{thm}

\section{Convergence of the lattice approximation for the interacting
theory -- in the sense of ultrahyperfunctions} For a
motivation of relativistic quantum field theory in terms of
tempered ultra-hyperfunctions as the appropriate framework
for a relativistic quantum field theory with a fundamental
length and for a brief introduction to the mathematics of
such a theory we have to refer to [\cite{BN04}]. Here we
just mention the basic definitions and results about
tempered ultra-hyperfunctions as we need them.

 For a subset $A$ of $\R^{n}$, we denote by $T(A)=\R^{n}+
 i A \subset  \C^{n}$ the tubular set with base $A$. For a convex compact
 set $K$ of $ \R^{n}$, ${\mathcal T}_{b}(T(K))$ is, by definition,
 the space of all continuous functions $f$ on $T(K)$ which are holomorphic in the
 interior of $T(K)$ and satisfy
$$
 \Vert f\Vert^{T(K), j}=\sup \{\vert z^{p}f(z)\vert ; z \in T(K), \vert p\vert
  \leq j\}<\infty,\  j = 0, 1,\ldots$$
where $p = (p_{1}, \ldots , p_{n})$ and $z^{p} = z_{1}^{p_{1}} \cdots z_{n}^{p_{n}}$.  ${\mathcal T}
_{b}(T(K))$ is a Fr\'echet space with the semi-norms $\Vert f\Vert ^{T(K), j}$.  If
$K_{1} \subset  K_{2}$ are two compact convex sets, we have the canonical
injections:
$$
{\mathcal T}_{b}(T(K_{2}))\rightarrow {\mathcal T}_{b}(T(K_{1})).
$$
Let $O$ be a convex open set in $\R^{n}$. We define
$$
{\mathcal T}(T(O))=\lim_{\leftarrow }{\mathcal T}_{b}(T(K_{1})),
$$
where $K_{1}$ runs through the convex compact sets contained in $O$ and
the projective limit is taken following the restriction mappings.
\begin{defn} \label{def4.1}
 A tempered ultra-hyperfunction is by definition a
continuous linear functional on ${\mathcal T} (T(\R^{n}))$.
\end{defn}
Characterizations of tempered ultra-hyperfunctions are known since
many years ([\cite{Ha61, Mo70, Mo75b}]). The most convenient one for
our purposes is based on a result in [\cite{BN04}] which we prepare
briefly.

Let ${\mathcal A}_{0}(W)$ be the space of all  functions
$F$, holomorphic in an open set $W\subset \C^{n}$, with the
property that for any positive numbers $\ep$, $K $, there
exist a multi-index $p$ and a constant $C \geq  0$ such
that
$$\vert F(z)\vert \leq  C(1+ \vert z^{p}\vert)\quad\textrm{for all} \,\
 z \in \C^{n} \backslash  (\C^{n} \backslash W)_{\epsilon},\;
 |{\rm Im\,}z_j|\leq K$$
where $(\C^{n} \backslash  W)_{\epsilon }$ is the open
$\epsilon -$neighbourhood of $(\C^{n} \backslash W)$. Let
$\sigma=(\sigma_1,\ldots,\sigma_n)$ be a vector with
components $\sigma_j \in \set{\pm 1} $. For such a vector
$\sigma$ and a number $k>0$ introduce the open set
$$\C^n_{\sigma,k}=\{(z_1,\ldots,z_n) \in \C^n; \sigma_j {\rm
Im\,}z_j > k,\;{\rm for\;} j=1,\ldots,n\}$$ and the space
$\A_0(\C^n_{\sigma,k})$ introduced above. Next we consider
collections $\set{F_{\sigma}}$ of elements $F_{\sigma} \in
\A_0(\C^n_{\sigma,k})$. Furthermore, for $\ep >0$, $k>0$,
and $\sigma_j \in \set{\pm 1}$, define the path
$$\Gamma_{\sigma_j}\equiv \Gamma_{\sigma_j}(\ep,k)\eqdef \set{z \in
\C; z=x+i\sigma_j(k+\ep), x \in \R}.
$$
and then the product path $\Gamma_{\sigma}= \prod_{j=1}^n
\Gamma_{\sigma_j}$.

Then from the definition of the spaces ${\mathcal T}
(T(\mathbb R^{n}))$ and ${\mathcal A} _{0}(\mathbb
C^{n}_{\sigma , k})$ it is clear that for any collection
$\set{F_{\sigma}}$ of $F_{\sigma } \in {\mathcal A}
_{0}(\mathbb C^{n}_{\sigma , k})$ the assignment
  \beq
\label{uhf-int-def} {\mathcal T} (T(\mathbb R^{n})) \ni  f
\rightarrow \langle \{ F_{\sigma }\}, f\rangle  = \sum
_{\sigma } \sigma _{1} \cdots \sigma _{n} \int _{\Gamma
_{\sigma }} F_{\sigma }(z) f(z) dz \in \mathbb C \eeq is
well defined and for fixed collection $\{ F_{\sigma }\} $
is linear and continuous in $f \in {\mathcal T} (T(\mathbb
R^{n}))$.  Thus for given collection $\{F_{\sigma}\}$,
$F_{\sigma} \in{\mathcal A} _{0}(\mathbb
C^{n}_{\sigma,k})$,
$${\mathcal T} (T(K))\ni f \rightarrow  \langle
\{ F_{\sigma}\} , f\rangle $$ is a tempered
ultra-hyperfunction.  Conversely, it is shown in
[\cite{Ha61, BN04}] that for any element $M$ of ${\mathcal
T} (T(\mathbb R^{n}))^{\prime }$, there exist constant
$k>0$ and a collection $\{ F_{\sigma }\} $ of functions
$F_{\sigma }$ in ${\mathcal A} _{0}(\mathbb C^{n}_{\sigma ,
k})$
such that \beq \label{uhf-int-def2} M(f)= \langle
\{F_{\sigma }\}, f\rangle  \eeq
 for all $f \in  {\mathcal T} (T(\mathbb R^{n}))$ (see also [\cite{Ha61, Mo70,
Mo75b}]). This proves
\begin{thm}[characterization tempered ultra-hyperfunctions]\label{thm:tuhf}
 A linear functional
$M$ on ${\mathcal T} (T(\R^{n}))$ is a tempered
ultra-hyperfunction if, and only if, it is of the form
(\ref{uhf-int-def}), (\ref{uhf-int-def2}) for some $k>0$
and some collection $\{ F_{\sigma }\} $ of functions
$F_{\sigma }$ in ${\mathcal A} _{0}(\mathbb C^{n}_{\sigma ,
k})$.
\end{thm}
\begin{rem}\label{rem:uhf-def}
 In  quantum field theory with a
fundamental length, often  functionals appear which are defined
for $g \in {\mathcal T} (T(\mathbb R^{2\cdot 4}))$ by \beq
\label{uhf-qft} \langle F, g\rangle=\int F(x_{1}^{0} - x_{2}^{0} -
i(k + \epsilon ), \mbox{\bi x}_{1} - \mbox{\bi x}_{2})g(x_{1}^{0}-
i(k + \epsilon ), \mbox{\bi x}_{1}, x_{2}^{0}, \mbox{\bi x}_{2})
dx_{1}^{0} \cdots dx_{2}^{3}$$
$$= \int F(x^{0} - i(k + \epsilon ), \mbox{\bi x})f(x^{0}-i(k +
\epsilon ), \mbox{\bi x}) dx^{0} \cdots dx^{3} \eeq for an
analytic function $F \in  {\mathcal A} _{0}(W)$ defined in
the region
$$      W = \{ z = (z^{0}, \ldots , z^{3}) \in  \mathbb C^{4}; - {\rm Im \, }z^{0} > \vert {\rm Im \, }\mbox{\bi z}\vert  + k\} $$
and
$$      f(z) = \int g(z + x_{2}, x_{2}) dx_{2}^{0} \cdots dx_{2}^{3} \in  {\mathcal T} (T(\mathbb R^{4})).$$
It is clear that the integral (\ref{uhf-qft}) defines a tempered
ultra-hyperfunction. But the integral representation
(\ref{uhf-qft}) looks quite different from the integral
representation (\ref{uhf-int-def}), (\ref{uhf-int-def2}) which
characterizes tempered ultra-hyperfunctions $M$ according to
Theorem \ref{thm:tuhf}.  Here we explain that the integral
representation (\ref{uhf-int-def}), (\ref{uhf-int-def2}) can be
expressed by the integral (\ref{uhf-qft}) in certain situations
(e.g., the support of the Fourier transformation $\tilde{M} $ of
$M$ is contained in the forward light-cone $\bar{V} _{+} = \{ x\in
\mathbb R^{4}; x^{0} \geq  \vert \mbox{\bi x}\vert \} $).  For
simplicity, we assume $n = 2$. Consider the situation that
$F_{(1,1)} = F_{(-1,1)} = 0$. Then
$$\langle F, f\rangle =\sum_{\sigma}\sigma_{1}\sigma_{2}
\int_{\Gamma_{\sigma}}F_{\sigma}(z) f(z) dz
 =\sum_{\sigma =(1,-1), (-1,-1)} \sigma_{1}\sigma_{2}\int_{\Gamma _{\sigma }} F_{\sigma }(z) f(z) dz$$
$$      = -\int \int F_{(1,-1)}(x_{1}+i(k+\epsilon ), x_{2}-i(k+\epsilon )) f(x_{1}+i(k+\epsilon ), x_{2}-i(k+\epsilon )) dx_{1}dx_{2}$$
$$      + \int \int F_{(-1,-1)}(x_{1}-i(k+\epsilon ), x_{2}-i(k+\epsilon )) f(x_{1}-i(k+\epsilon ), x_{2}-i(k+\epsilon )) dx_{1}dx_{2}.$$
Now we further assume that $F_{(\pm 1,-1)}$ is analytically continued
from
$$      \mathbb C^{2}_{(\pm 1,-1),k} = \{ (z_{1}, z_{2}) \in  \mathbb C^{2}; \pm  {\rm Im \, }z_{1} > k, \  - {\rm Im \, }z_{2} > k\}  $$
to the set
$$      \{ (z_{1}, z_{2}) \in  \mathbb C^{2}; \pm  {\rm Im \, }z_{1} - {\rm Im \, }z_{2} > 2k, \  - {\rm Im \, }z_{2} > k\} .$$
Then by deforming the path of integration, we get
$$      \int \int F_{(1,-1)}(x_{1}+i(k+\epsilon ), x_{2}-i(k+\epsilon )) f(x_{1}+i(k+\epsilon ), x_{2}-i(k+\epsilon )) dx_{1}dx_{2}$$
$$      = \int \int F_{(1,-1)}(x_{1}, x_{2}-i(2k+\epsilon )) f(x_{1}, x_{2}-i(2k+\epsilon )) dx_{1}dx_{2}$$
and
$$\int\int F_{(-1,-1)}(x_{1}-i(k+\epsilon),x_{2}-i(k+\epsilon))f(x_
{1}-i(k+\epsilon ), x_{2}-i(k+\epsilon )) dx_{1}dx_{2}$$
$$= \int \int F_{(-1,1)}(x_{1}, x_{2}-i(2k+\epsilon)) f(x_{1},
x_{2}-i(2k+\epsilon )) dx_{1}dx_{2}.$$ Put $G(z_{1}, z_{2}) =
-F_{(1,-1)}(z_{1}, z_{2}) - F_{(-1,1)}(z_{1}, z_{2})$.  Then
$G(z_{1}, z_{2})$ is analytic in
$$\{(z_{1},z_{2}) \in\mathbb C^{2};({\rm Im\,}z_{1}-{\rm Im,}
z_{2}>2k)\wedge(-{\rm Im\,}z_{1}-{\rm Im\,}z_{2}> 2k)\}$$
$$=\{(z_{1}, z_{2}) \in  \mathbb C^{2}; - {\rm Im \, }z_{2} >
\vert {\rm Im \, }z_{1}\vert  + 2k\}, $$ and we have
$$\langle F, f\rangle=\int \int G(x_{1},x_{2}+i(2k+\epsilon))
f(x_{1}, x_{2}+i(2k+\epsilon))dx_{1}dx_{2}.$$
\end{rem}
\medskip

As we had seen at the end of Section 2, the $n$-point
Wightman function of the field $\psi(x)$
is
\beq \label{W-functions}
{\mathcal W}^{r}_{\alpha }(x_{1}, \ldots , x_{n})= (\det A)^{-1/2} {\mathcal W} ^{r}_{0,\alpha }(x_{1}, \ldots , x_{n}), \eeq
where $A$ is the matrix determined by (\ref{abbrev2}), i.e.,
$(a_{j,k}),\;j,k=1,\cdots,n$ with
$$      a_{j,j} = 1, \  a_{j,k} = a_{k,j} = 2h_{r_{j}}h_{r_{k}}l^{2}
D_{m}^{(-)}(x_{j} - x_{k}), {\rm if}\; j< k.$$

In [\cite{BN04}] the functional characterization of a relativistic quantum field theory
with a fundamental length has been given in terms of six conditions (R0) $\cdots$ (R5). Now we are going to show that the system (\ref{W-functions}) satisfies condition (R0) which states that this systems consists of symmetric tempered ultra-hyperfunctions. The first part of this condition (R0) says that the assignment
$${\mathcal T} (T(\R^{4n}) \ni  f \rightarrow
{\mathcal W}^{r_{1}, \ldots , r_{n}}_{\alpha_{1}, \ldots , \alpha_{n}}(f) \in  \C$$
is a continuous linear functional on ${\mathcal T} (T(\R^{4n}))$, for $n=1,2,3,\ldots$.

In order to investigate this continuity property,  we apply the general expansion formula for determinants we get
$$\det A =\sum {\rm sgn\,}(j, k,\ldots, l)a_{1,j}a_{2,k}
\cdots a_{n, l}$$
$$=a_{1,1}a_{2,2}\cdots a_{n,n}+\sum_{(j,k,\ldots,l)\neq(1,2,
\ldots,n)}{\rm sgn\,}(j,k,\ldots,l)a_{1,j}a_{2,k}\cdots a_{n,l}.$$
Because of the special values of the entries $a_{j,k}$ according to
(\ref{abbrev2}) we see
\begin{equation}\label{det-A}
\det{A} =1 + P_n(a_{j,k})
\end{equation}
where $P_n(a_{j,k})$ is the sum of homogeneous polynomials of
degrees $m=2,\cdots,n$ in the entries $a_{j,k},\;1\leq j <k \leq
n$ with integer coefficients.
The integral representation for $D_{m}^{(-)}$ as given at the end of
Section 3 easily implies, for every $\epsilon >0$, the global
estimate
\begin{equation}\label{estim-2point}
 \vert D_{m}^{(-)}(x^{0}-i\epsilon,\mbox{\bi x})\vert \leq (2\pi
\epsilon)^{-2} \quad \textrm{for all}\;x \in \R^{4}.
\end{equation}
 It follows that $\vert P(a_{j,k})\vert < 1$
 if we choose all $y_{k}^{0} - y_{j}^{0},\;j<k$, sufficiently large
 and put $z_{j} =(x_{j}^{0} + iy_{j}^{0}, \mbox{\bi x}_{j})$.
 Hence for these $z_{j}$,
$(\det A(z))^{-1/2} = (1 + P(a_{j,k}(z_{j}, z_{k})))^{-1/2}$ is a
bounded analytic function of the $x_{j}$ in a tubular domain and
therefore, according to Theorem \ref{thm:tuhf},  $${\mathcal
W}^{r}_{\alpha}(z_{1},\ldots,z_{n}) = (\det A(z))^{-1/2}{\mathcal
W}^{r}_{0,\alpha}(z_{1},\ldots,z_{n})$$ determines a tempered
ultra-hyperfunction by the formula
\begin{equation}\label{eq:n-point}
 {\mathcal
W}^{r}_{\alpha}(f)=\int_{\prod_{j=1}^{n}\Gamma_{j}}(\det
A(z))^{-1/2} {\mathcal W}^{r}_{0,\alpha}(z_{1},\ldots,z_{n}) f(z)
dz,
\end{equation}
where $\Gamma_{j}=\R^{4} + i(y_{j}^{0},0,0,0)$, for all $f \in
{\mathcal T}(T(\R^{4n}))$, i.e., the first part of Condition (R0) is satisfied.

We conclude that the sequence of Wightman functions ${\mathcal
W}^{r}_{\alpha}$ satisfies
${\mathcal W}^{r}_{\alpha} \in {\mathcal T}(T(\R^{4n}))\pr$ for $n=1,2,3, \ldots$ .
The second part of Condition (R0), i.e.,
$$  {\mathcal W}^{\bar{r}_n, \ldots , \bar{r}_1}_{\bar{\alpha}_n, \ldots , \bar{\alpha}_1}(f^{*}) = \overline{{\mathcal
W}^{r_1, \ldots , r_n}_{\alpha _1, \ldots , \alpha _n}(f)}, \
f^{*}(z_1, \ldots , z_n) = \overline{f(\overline{z}_n, \ldots ,
\overline{z}_1)},
$$ where $\psi ^{r*} _{\alpha} = \psi ^{\bar{r}}
_{\bar{\alpha}}$, follows easily from the fact that
$$  \overline{D_{m}^{(-)}(z_{j}-z_{k})} =
D_{m}^{(-)}(\overline{z}_{k}-\overline{z}_{j}). $$

To conclude this section, we have a closer look at the two-point function
$${\mathcal W}^{1, 2}_{\alpha_{1}, \alpha_{2}}(z_{1}, z_{2}) = [1 - 4l^{4}D_{m}(z_{1} - z_{2})^{2}]^{-1/2} {\mathcal W}^{1, 2}_{0, \alpha _{1}, \alpha _{2}}(z_{1}, z_{2}).$$
Estimate (5.6) shows that $\vert 4l^{4}D_{m}(x^{0}_{1} - x^{0}_{2} - i\epsilon , \mbox{\bi x}_{1} - \mbox{\bi x}_{2})^{2}\vert  < 1$
if $\epsilon  > \ell  = l/(\sqrt{2}\pi )$, and $[1 - 4l^{4}D_{m}(x^{0}_{1} - x^{0}_{2} - i\epsilon , \mbox{\bi x}_{1} - \mbox{\bi x}_{2})^{2}]^{-1/2}$
is analytic with respect to $x_{1}$ and $x_{2}$.  Then the functional
defined by (5.7) for $n = 2$ and $y_{2} - y_{1} = \epsilon  > 0$ can distinguish
the two events only if their distance is greater than $\epsilon $ (see
[\cite{BN04}]).  Since $\epsilon  > \ell $ is arbitrary, $\ell $ is the fundamental
length of our theory.

\section{Conclusion}
The results of this article provide a solution of the linearized version of Heisenberg's
fundamental equation, on the level of  all the $n$-point functionals of the solution fields.
This has been achieved by employing path integral methods for quantization. In order to have
all the path integrals well defined and to evaluate them rigorously, a lattice approximation was used and the continuum limit
of this approximation was controlled by using non-standard analysis. This continuum limit exists
in the framework of tempered ultrahyperfunctions but not in the framework of tempered
distributions. In this way in particular the convergence of the lattice approximations
for a free scalar field, a free Dirac field and for the interacting fields of this model
has been established.

In the second part we are going to show that the sequence of all $n$-point functionals
which we have constructed satisfy all the defining conditions of a relativistic quantum
field theory with a fundamental length, in the sense of [\cite{BN04}]. We do so by first
extending the theory of [\cite{BN04}] to include scalar as well as  spinor fields
and then verifying the defining condition. In addition we offer an alternative way
to calculate all the $n$-point functionals of the theory by use of Wick power series
which converge in the sense of tempered ultrahyperfunctions. And its is shown that the
solution fields $(\phi,\psi)$ of (\ref{Oku-eq1}) - (\ref{Oku-eq2})
can be express of a point-wise product
\beq \label{eq:solution}
\psi(x) = \psi_0(x) :e^{il^2 \phi(x)^2}:
\eeq

where $\psi_0$ is the free Dirac field.\\[0.5cm]

{\bf Acknowledgements}. This paper is part of the research project {\it
Research for axiomatic quantum field theory by using ultrahyperfunctions},
grant number 16540159, Grants-in-Aid for Scientific Research of JSPS (Japan Society for the
Promotions of Science). The authors gratefully acknowledge substantial financial support by the
JSPS.

Major parts of the work for this article were done
during a research visit of E. B. to the University of Tokushima, funded by this grant. With great pleasure,
E. B. expresses his gratitude to the Department of Mathematics, in particular his host
S. N., and the JSPS.
\bibliographystyle{plain}
\bibliography{hfqft20071224}

\end{document}